\documentclass[12pt]{article}
\usepackage{hepth}

\newcommand{\oss}[1]{{(\overline{{#1}})}}
\newcommand{\pdt}{\left(\substack{\hbox{\tiny possible} \\ \hbox{\tiny divergent} \\ \hbox{\tiny terms}}\right)}
\newcommand{\rhos}{\rho_{\rm s}}
\newcommand{\rhot}{\rho_{\rm t}}
\newcommand{\rhon}{\rho_{\rm n}}

\def\be{\begin{equation}}
\def\ee{\end{equation}}
\def\myop{{\langle O_\psi \rangle}}

\title{Transport in holographic superfluids}

\authors{Christopher P. Herzog\worksat{\PU,}\footnote{e-mail: {\tt cpherzog@princeton.edu}},
Nir Lisker\worksat{\Technion,}\footnote{e-mail: {\tt snlisker@t2.technion.ac.il}},
Piotr Sur\'owka\worksat{\VUB}\footnote{e-mail: {\tt piotr.surowka@vub.ac.be}},
and
\\
Amos Yarom\worksat{\PU,\Technion,}\footnote{e-mail: {\tt ayarom@princeton.edu}}}

\institution{PU}{Joseph Henry Laboratories, Princeton University, Princeton, NJ 08544}
\institution{Technion}{Department of Physics, Technion, Haifa 32000, Israel}
\institution{VUB}{Theoretische Natuurkunde, Vrije Universiteit Brussel, Pleinlaan 2, B-1050 Brussels, Belgium}

\abstract{We construct a slowly varying space-time dependent holographic superfluid and compute its transport coefficients.  Our solution is presented as a series expansion in inverse powers of the charge of the order parameter. We find that the shear viscosity associated with the motion of the condensate vanishes. The diffusion coefficient of the superfluid is continuous across the phase transition while its third bulk viscosity is found to diverge at the critical temperature. As was previously shown, the ratio of the shear viscosity of the normal component to the entropy density is $1/4\pi$. As a consequence of our analysis we obtain an analytic expression for the backreacted metric near the phase transition for a particular type of holographic superfluid.}

\preprint{PUPT-2363}

\begin{document}

\maketitle

\tableofcontents

\section{Introduction and summary}
\label{S:Introduction}

It is quite remarkable that black holes in asymptotically AdS spacetimes can be applied, through the AdS/CFT correspondence \cite{Maldacena:1997re,Gubser:1998bc,Witten:1998qj}, to myriad physical systems. Among these systems are gauge-theory plasmas whose hydrodynamic behavior is associated with the topography of the black hole horizon \cite{Bhattacharyya:2008jc}, and superfluid states of large $N$ gauge-theories \cite{Hartnoll:2008vx,Herzog:2008he,Basu:2008st}. In this work we show how, in the presence of a charged scalar condensate, the black hole horizon encodes superfluid properties of the dual gauge theory and compute various generic properties of the superfluid.

The membrane paradigm \cite{MembraneParadigm} suggests a relationship between fluid dynamics and black holes.  In the context of AdS/CFT, ref.\ \cite{Policastro:2001yc} made this relation precise by computing the ratio of the shear viscosity, $\eta$, to entropy density, $s$, of $\mathcal{N}=4$ Super-Yang Mills
\begin{equation}
\label{E:Universal}
	\frac{\eta}{s} = \frac{1}{4\pi}
\end{equation}
using a Kubo formula. The ratio in \eqref{E:Universal} is considered to be universal in the sense that it is valid,  in the supergravity limit, for a large class of theories \cite{Kovtun:2004de,Buchel:2003tz,Buchel:2004qq,Brustein:2009rn}.\footnote{See \cite{Erdmenger:2010xm} for a recent example where the authors argue that \eqref{E:Universal} is, in some sense, violated in a non-isotropic holographic superfluid.} The work of \cite{Bhattacharyya:2008jc} provided a new layer of development in describing fluid flow using AdS/CFT by demonstrating a one-to-one correspondence between solutions of relativistic Navier-Stokes equations in field theory and solutions of Einstein's equations in the dual black hole geometry.
Indeed, using the formalism of \cite{Bhattacharyya:2008jc} new transport coefficients were unveiled \cite{Erdmenger:2008rm,Banerjee:2008th,Torabian:2009qk,Son:2009tf,Eling:2010hu,Haack:2008xx} and relations similar to \eqref{E:Universal} were found for higher order transport coefficients \cite{Haack:2008xx}.

Superfluidity, initially discovered in liquid helium, can be thought of as a fluid with a spontaneously broken global symmetry. At temperatures below $T_0 \sim 2.17^{\circ}\;K$ liquid helium undergoes a phase transition into a superfluid state. Many of the remarkable properties of superfluid helium can be attributed to the lack of viscosity of its superfluid component. In \cite{Tisza} and later in \cite{Landau} Tisza and Landau formulated a hydrodynamic model in which a superfluid is described by a two component fluid: a condensate which, roughly speaking, has no viscosity and a normal component which is viscous.

As we review in section \ref{S:Superfluids}, one can extend the Tisza-Landau model to relativistic superfluids. 
Several relativistic versions of the inviscid superfluid can be found in the literature \cite{Khalatnikov198270,PhysRevD.45.4536,Carter1992243,Israel198179,Israel198277,Son:2000ht}. 
Viscous corrections to these models have been treated, to a certain extent, in \cite{Pujol:2002na,Valle:2007xx,Gusakov:2006ga,Gusakov:2007px} and applied to studies of neutron stars and cold relativistic superfluids in \cite{Gusakov:2007px,Mannarelli:2009ia}.
The number of transport coefficients associated with viscous corrections can be quite large.
 For example,
the author of \cite{Putterman} counts 13 transport coefficients in the non-relativistic theory when the relative superfluid velocity is not small. 

One can make several assumptions that reduce the number of transport coefficients to three.
Because superfluidity is typically lost if the relative velocity between the superfluid and normal components becomes too great, a common assumption 
is that this relative velocity is small, $n^\mu \ll 1$.  Another common assumption is the absence of parity breaking terms in the hydrodynamic expansion.
In view of our applications to holographic superfluids with a traceless stress-energy tensor, we also assume conformal invariance. 
At leading order in the hydrodynamic gradient expansion, 
these assumptions reduce the number of transport coefficients to four: the shear viscosity $\eta$, 
a shear viscosity associated with the superfluid component $\eta_s$, a diffusivity $\kappa$ associated with charge transport or heat conductivity, and a superfluid bulk viscosity
$\zeta_3$.  (The bulk viscosities conventionally called $\zeta_1$ and $\zeta_2$ are eliminated by the assumption of conformal invariance.)
With an additional assumption on the form of the entropy current, one can argue that $\eta_s=0$.  
Thus, we are left with the three transport coefficients $\eta$, $\kappa$, and $\zeta_3$. 

A holographic superfluid is a superfluid with a dual higher dimensional gravitational description.
In gravity, and in the simplest possible setup, such configurations involve a charged scalar field and a $U(1)$ gauge field, both coupled to the metric \cite{Gubser:2008px,Hartnoll:2008vx}.\footnote{More elaborate configurations dual to $p$-wave and $d$-wave superfluids can be found in \cite{Gubser:2008wv,Chen:2010mk,Benini:2010pr}.} Various aspects of these relativistic superfluid configurations were studied in \cite{Herzog:2008he,Basu:2008st,Yarom:2009uq,Herzog:2009md,Amado:2009ts,Gubser:2009qf,Keranen:2009ss,Keranen:2009re,Sonner:2010yx,Arean:2010wu,Keranen:2010sx}.

In this work we construct a solution to the equations of motion that follow from the classical gravity system dual to a space-time dependent superfluid and use it to compute the transport coefficients described above. The challenge in extending the work of \cite{Bhattacharyya:2008jc} to the holographic superfluids of \cite{Hartnoll:2008vx} is that \cite{Bhattacharyya:2008jc} relies on analytic techniques whereas the holographic superfluid is usually constructed numerically. To simplify our analysis, we work in a configuration in which the charge of the scalar, $q$, is large. We go beyond the probe limit of \cite{Hartnoll:2008vx}, allowing the metric to backreact on the matter fields. Within our approximation we can establish analytically that the shear viscosity of the superfluid component $\eta_s$ vanishes, the diffusion coefficient $\kappa$ is continuous across the phase transition and that close to the phase transition the bulk viscosity $\zeta_3$ scales like the inverse order parameter (and hence diverges). For the particular analytic solution of \cite{Herzog:2010vz} we are able to obtain explicit expressions for  the third bulk viscosity and the diffusion coefficient near the phase transition which occurs at $\mu /T = 2\pi/q$:
\begin{align}
\label{E:FinalKappa}
	\kappa&= \frac{T^2}{\kappa_5^2} \left( \frac{\pi}{2} - 3 \left( \frac{q\mu}{T} - 2\pi \right) \right) +\mathcal{O} \left( \left( \frac{q\mu}{T} - 2\pi  \right)^2 \right)  \\
\label{E:FinalZeta3}
	\zeta_3 & =\frac{\kappa_5^2}{T^3} \left( \frac{13}{294 \pi^2} \left( \frac{q\mu}{T} - 2\pi  \right)^{-1} + \frac{21821-37152 \ln 2}{49392 \pi^3}\right) + \mathcal{O}  \left( \frac{q\mu}{T} - 2\pi  \right) 
\end{align}
where $\kappa_5^2$ is associated with Newton's constant in $AdS_5$. 
It is usually related to the rank of the gauge group $N_c$ through a relation of the form $\kappa_5^2 \propto N_c^{-2}$.
The peculiar dependence of $\zeta_3$ on $\kappa_5^2$ will be discussed in section \ref{S:Discussion}.

The remainder of this work is organized as follows. In section \ref{S:Superfluids} we derive the constitutive relations for relativistic superfluids, generalizing the work of \cite{Son:2000ht,Pujol:2002na,Valle:2007xx,Gusakov:2007px}. 
Section \ref{S:method} provides an overview of the computations which appear in sections \ref{S:Static} and \ref{S:Dynamical}.
In section \ref{S:Static} we discuss the large charge holographic superfluid. The hydrodynamic fields discussed in section \ref{S:Static} are not spacetime dependent and, with some abuse of language, we refer to these configurations as static. Most of the material in the latter section is contained in \cite{Hartnoll:2008vx,Herzog:2008he,Herzog:2010vz} but we alert the interested reader to several new features of the static superfluid which have not been discussed elswhere. These include an extension of the solution of \cite{Herzog:2010vz} to include backreaction of the metric and an analytic derivation of the Josephson condition which was observed numerically in \cite{Sonner:2010yx}. In section \ref{S:Dynamical} we use a gradient expansion to extend the static holographic superfluid solution to a dynamical one and use it to show that $\zeta_3$ diverges near the phase transition in section \ref{S:GEPO0}, that $\kappa$ is continuous in section \ref{S:GEPO1} and that $\eta/s=1/4\pi$ and $\eta_s=0$ in section \ref{S:GEPO2}. Equations \eqref{E:FinalKappa} and \eqref{E:FinalZeta3} together with corrections to the order parameter are derived in section \ref{S:ExplicitO1}. We end with a  discussion of our results in section \ref{S:Discussion}.
Appendix  \ref{A:BtoB} collects some useful formulae used in the latter sections of the paper while
appendix \ref{A:Kubo} provides an alternate derivation of the viscous transport coefficients by computing two-point functions from gravity and employing Kubo formulae. Some of the results of section \ref{S:Static} are gathered in appendix \ref{A:exactsols}.

While this work was nearing completion we learned of \cite{Minwalla} which has some overlap with the content of this paper.

\section{Relativistic superfluids}
\label{S:Superfluids}

We study a conformally invariant, relativistic superfluid in the limit where the relative superfluid velocity is small and there are no parity breaking terms in the hydrodynamic expansion.
While our main result for the constitutive relations of the energy momentum tensor and current can be found in the literature, we feel that the our construction provides a more transparent view of the physical assumptions used. In section \ref{S:Thermodynamics} we discuss the thermodynamic variables which describe the superfluid. We then use these variables in section \ref{S:Inviscid} to construct the energy momentum tensor and charged current of a superfluid in the inviscid limit. In section \ref{S:Viscous} we extend our analysis to include viscous corrections. In section \ref{S:Kubo}, we describe how Kubo relations for viscous transport coefficients can be extracted from linearized hydrodynamics, and in \ref{S:Sound} we investigate the speed and attenuation of first, second, and fourth sound in a relativistic, conformally invariant superfluid.

\subsection{Superfluid thermodynamics}

\label{S:Thermodynamics}
As discussed in section \ref{S:Introduction} a relativistic superfluid can be thought of as a fluid charged under a spontaneously broken $U(1)$ symmetry. In the grand canonical ensemble the thermodynamic variables characterizing normal, charged fluids are the temperature and chemical potential, $T$ and $\mu$, and their conjugate variables entropy and total charge density, $s$ and $\rhot$. Once the $U(1)$ symmetry is spontaneously broken then the goldstone boson $\phi$ provides for another degree of freedom. In equilibrium, the only extra gauge-invariant scalar degree of freedom we can construct from $\phi$ is $\partial_{\mu} \phi \partial^{\mu} \phi$ \cite{Herzog:2008he}.\footnote{%
 Here, Greek indices run from $0$ to $3$ and are raised 
 and lowered with the Minkowski metric $\eta_{\mu\nu} = (-+++)$.
}
 Thus, the thermodynamic variables used to describe the superfluid can be taken to be the magnitude of the gradient of the goldstone boson, the chemical potential, and the temperature:
$\partial_{\mu}\phi \partial^{\mu}\phi$, $\mu$ and $T$. As it turns out, the chemical potential and $\partial_{\mu}\phi \partial^{\mu}\phi$ are not independent. To see this we can turn on an external gauge field $A_{\mu}^{\hbox{\tiny ext}}$ whose time component couples to the charged current $J_{\mu}$ the same way a chemical potential does, 
\begin{equation}
\label{E:coupling}
	\mathcal{L} \to \mathcal{L} + A^{\hbox{\tiny ext}}_{\mu}J^{\mu}\,.
\end{equation} 
Replacing $\partial_{\mu}\phi$ with the gauge invariant combination $D_{\mu}\phi = \partial_{\mu}\phi - A_{\mu}^{\hbox{\tiny ext}}$ we conclude that $D_0\phi = -\mu$.\footnote{This relation is also called the Josephson condition. We will see it reappear when we discuss the hydrodynamics of superfluids.}
When $A_{\mu}^{\hbox{\tiny ext}}= 0$, the independent degrees of freedom are: $T$, $\mu$ and $\partial_i\phi\partial^i\phi = \partial_{\mu}\phi\partial^{\mu}\phi + \mu^2$ where $i=1,\ldots,3$ denote spatial directions.

We define the energy density of the system $\epsilon (s,\,\rhot,\,\partial_i\phi\partial^i\phi)$ as the time-time component of the stress tensor in the rest frame of the normal component. The pressure of the system is usually defined as the force per unit area in an equilibrium configuration. Since the superfluid velocity need not vanish in the rest frame of the normal component an equilibrium configuration is not well defined. Thus, we define the pressure $P$ through the relation
\begin{equation}
\label{E:gd}
	\epsilon = -P + s T + \mu \rhot
\end{equation}
and therefore we have
\begin{equation}
\label{E:dP}
	dP = s dT + \rhot d\mu + \frac{X}{2} d \left( \partial_{\mu}\phi \partial^{\mu}\phi- \mu^2  \right)\,.
\end{equation}
In \eqref{E:dP}, $X$ is defined to be the variable conjugate to $(\partial_{\mu} \phi)^2 - \mu^2$. We will see shortly that it is associated with the superfluid density. Our conventions here are slightly different from the ones used in \cite{Son:2000ht,Herzog:2008he,Gusakov:2006ga}. 

\subsection{Inviscid superfluid hydrodynamics}
\label{S:Inviscid}

Consider ideal superfluid hydrodynamics where the superfluid is described by a temperature $T$, a chemical potential $\mu$, a velocity field $u^{\mu}$ normalized such that $u^{\mu}u_{\mu}=-1$ and the vector quantity $\partial_{\mu}\phi$ 
all considered to be slowly varying functions of the space-time coordinates. In what follows we will choose $u^{\mu}$ to describe the velocity field of the uncondensed phase and $\partial^{\mu}\phi$ to describe the non-normalized velocity of the condensate. 
We define the energy density $\epsilon$ and charge density $\rhot$ as the time-time component of the stress tensor $T_{\mu\nu}$ and the time component of the charged current $J_\mu$ when in the rest frame of the normal component:
\begin{subequations}
\label{E:Defs}
\begin{align}
\label{E:Defenergy}
	u^{\mu}u^{\nu} T_{\mu\nu} &= \epsilon \\
\label{E:Defcharge}
	u^{\mu}J_{\mu} &= -\rhot\,.
\end{align}
\end{subequations}
It is standard to decompose the energy momentum tensor and current such that
\begin{align}
\begin{split}
\label{E:GeneralDecomposition}
T_{\mu\nu} &= \epsilon u_{\mu}u_{\nu} +  2 j^e_{(\mu}u_{\nu)}+ \pi_{\mu\nu} \\
	J_{\mu} & = \rhot u_{\mu} + j^c_{\mu}
\end{split}
\end{align}
where $j^e_{\mu} u^{\mu}  = j^c_{\mu}u^{\mu} = 0$ and $\pi_{\mu\nu}u^{\nu}=0$. 
Subscripted parentheses denote a symmetrized quantity
\begin{equation}
\label{E:SymmProduct}
	A_{(\mu\nu)} = \frac{1}{2}(A_{\mu\nu}+A_{\nu\mu})\,.
\end{equation} 
The vectors $j^e_{\mu}$ and $j^c_{\mu}$ describe the energy flux and charge flux in the $u^\mu$ frame respectively. At the inviscid level, the only vector orthogonal to the velocity field 
is $P^{\mu\nu}\partial_{\nu}\phi$ where $P_{\mu\nu}$ is a projection matrix orthogonal to the velocity field
\begin{equation}
\label{E:Projectiondef}
	P_{\mu\nu} = \eta_{\mu\nu} + u_{\mu}u_{\nu} \ .
\end{equation}
Since we will be using the combination 
$P_{\mu}^{\phantom{\mu}\nu}\partial_{\nu}\phi = \partial_{\mu}\phi + u_{\mu} u^{\nu}\partial_{\nu}\phi $ often, 
we find it useful to define two quantities
\begin{equation}
	\mu_{\rm s} \equiv -u^{\nu}\partial_{\nu}\phi \ , \; \; \; 
	\mu_{\rm s} n_{\mu} \equiv P_{\mu}^{\nu}\partial_{\nu}\phi\,.
\end{equation}
Note that
\begin{equation}
	(\partial_{\nu}\phi)(\partial^{\nu}\phi) = \mu_{\rm s}^2 \left(n^2-1\right)\,.
\end{equation}
Since $\partial_{\mu}\phi$ is proportional to the superfluid velocity, we interpret $n^{\mu}$ as the relative superfluid velocity. From the discussion in section \ref{S:Thermodynamics} we expect that $\mu_{\rm s}= \mu$. We will soon verify this expectation. 
 
Following \eqref{E:GeneralDecomposition} we can write the charged current as
\begin{align}
\begin{split}
\label{E:JAnsatz}
	J^{\mu} & = \rhot u^{\mu} + \tau_1 \mu_{\rm s} n^{\mu} \\
		&=\left(\rhot-\tau_1 \mu_{\rm s}\right)u^{\mu} + \tau_1 \partial^{\mu}\phi
\end{split}
\end{align}
where $\tau_1 = \tau_1(\mu,T, \partial_\mu\phi P^{\mu\nu}\partial_{\nu}\phi)$.
If we assume that $u_{\mu}$ carries the charge of the normal component then we should make the identification 
\begin{equation}
\label{E:tau1}
	\rhos = \tau_1 \mu_{\rm s}
\end{equation}
where $\rhon = \rhot-\rhos$ is the charge carried by the normal phase.
Because of length contraction, in a relativistic setting there is no invariant notion of charge density.
When we specify the charge density, we must also specify the frame.   The charge density $\rhos$ defined above is the superfluid density in the rest frame of the normal component.  In the frame where the condensate is at rest, the charge density $\overline \rhos$ is smaller: $\overline \rhos = \rhos  \mu_{\rm s} / |\partial \phi|$.

The most general expression we can write for $T_{\mu\nu}$ is
\begin{equation}
\label{E:TmnAnsatz}
	T_{\mu\nu} = \epsilon u_{\mu}u_{\nu} + P_{\mu\nu}\left(P + \tau_2 \mu_{\rm s}^2 n^2\right) + 2 \tau_3 \mu_{\rm s} n_{(\mu}u_{\nu)} + \tau_4 \mu_{\rm s}^2 n_{\mu} n_{\nu}
\end{equation}
where $P$ was defined in \eqref{E:gd}, 
and $\tau_a = \tau_a(\mu,T,  \partial_\mu\phi P^{\mu\nu}\partial_{\nu}\phi)$, $a=2,3,4$. 
The coefficient of $P_{\mu\nu}$ is fixed by the requirement that when the relative superfluid velocity vanishes $n^{\mu}=0$, expression \eqref{E:TmnAnsatz} reduces to the familiar energy momentum tensor of a normal fluid.

To determine $X$ (defined in \eqref{E:dP}), the $\tau_a$, and $\mu_{\rm s}$, consider the entropy current $J_{s}^{\mu}$ which is assumed to be carried only by the normal component
\begin{equation}
\label{E:SCurrent}
	J^{\mu}_{s} = s u^{\mu}\,,
\end{equation}
and is conserved in the absence of viscous terms. Entropy conservation together with \eqref{E:gd} and
\eqref{E:dP} allow one to reduce
\begin{equation}
\label{E:useful}
	u_{\mu} \partial_{\nu} T^{\mu\nu} + \mu \partial_{\mu}J^{\mu} = 0\,
\end{equation}
into a constraint on $X$, the $\tau_a$, and $\mu_{\rm s}$. Explicitly, one finds
\begin{align}
\begin{split}
\label{E:InviscidFinal}
	\mu \partial_{\nu} J^{\nu} + u_{\mu} \partial_{\nu} T^{\mu\nu} =&
	-T \partial_{\alpha} (s u^{\alpha}) + \frac{1}{2} u_\nu \partial^\nu (\mu_{\rm s}^2 n^2) (X + \tau_4) \\
	&+ u_{\mu} u_\nu \partial^\nu (\mu_{\rm s} n^{\mu}) \left(\tau_3 - \mu_{\rm s} \tau_4\right) \\
	&+\mu_{\rm s} n^{\nu} \left(-\partial_{\nu} \tau_3 + \tau_4 \partial_{\nu} \mu_{\rm s} + \mu \partial_{\nu} \frac{\rho_{\rm s}}{\mu_{\rm s}} \right) \\
	&+\partial_{\nu} (\mu_{\rm s} n^{\nu} ) \left(-\tau_3 + \frac{\mu \rho_{\rm s}}{\mu_{\rm s}} \right) 
	- \tau_2 \mu_{\rm s}^2 n^2 \partial_{\alpha} u^{\alpha} 
\end{split}
\end{align}
which implies
\begin{equation}
\label{E:InviscidParameters}
	\tau_2 = 0 \qquad
	\tau_3 = \frac{\mu \rhos}{\mu_{\rm s}}   \qquad
	\tau_4 = \frac{\mu \rhos}{\mu_{\rm s}^2} \qquad
	X = -\frac{\mu \rhos}{\mu_{\rm s}^2} \,.
\end{equation}
and $\mu_{\rm s} \propto \mu$. In what follows we make the choice 
\begin{equation}
\label{E:JosephsonIni}
	\mu_{\rm s} = \mu\,.
\end{equation}
This freedom in choosing the ratio between $\mu_{\rm s}  $ and $\mu$ is due to a freedom of the overall normalization of the goldstone boson. The relation \eqref{E:JosephsonIni} is known as the Josephson condition.

To summarize, entropy conservation together with the requirement that the superfluid component carries no entropy implies that at the inviscid level
\begin{subequations}
\label{E:InviscidHydro}
\begin{align}
\label{E:InviscidTmn}
	T_{\mu\nu} &= (\epsilon+P ) u_{\mu}u_{\nu} + P \eta_{\mu\nu} + 2 \rhos u_{(\mu} P_{\nu)}^{\phantom{\mu}\alpha} \partial_{\alpha}\phi  +\frac{\rhos}{\mu}  P_{\mu}^{\phantom{\mu}\beta} P_{\nu}^{\phantom{\nu}\alpha} \partial_{\beta}\phi \partial_{\alpha}\phi  \\
\label{E:InviscidJ}
	J_{\mu} &= \rhot u_{\mu} + \frac{\rhos}{\mu} P_{\mu}^{\phantom{\mu}\alpha} \partial_{\alpha}\phi 
	\\
\label{E:InviscidJosephson}
	u^{\mu} \partial_{\mu} \phi &= -\mu\,.
\end{align}
\end{subequations}
The thermodynamic relations are given by
\begin{align}
\begin{split}
\label{E:Thermo}
	\epsilon+P &= s T + \mu \rhot \\
	dP &=  s dT + \rhon d\mu - \frac{\rhos}{2 \mu} d \left( \partial_{\nu}\phi \partial^{\nu}\phi \right) \\
		& = s dT + \rhot d\mu - \frac{\rhos}{2\mu} d (\partial_{\nu}\phi \partial^{\nu}\phi + \mu^2)
\end{split}
\end{align}

\subsection{Viscous superfluid hydrodynamics}
\label{S:Viscous}

With the definitions \eqref{E:Defs} and the inviscid solution \eqref{E:InviscidHydro} we can parametrize the viscous corrections to the energy momentum tensor, current and Josephson condition by $V_{\mu}^e$, $\tau_{\mu\nu}$, $\Upsilon_{\mu}$ and $\upsilon$,
\begin{subequations}
\label{E:ViscousHydro1}
\begin{align}
	T_{\mu\nu} &= (\epsilon+P ) u_{\mu}u_{\nu} + P \eta_{\mu\nu} + 2 \rhos u_{(\mu} P_{\nu)}^{\phantom{\mu}\alpha} \partial_{\alpha}\phi  +\frac{\rhos}{\mu}  P_{\mu}^{\phantom{\mu}\beta} P_{\nu}^{\phantom{\nu}\alpha} \partial_{\beta}\phi \partial_{\alpha}\phi  + 2 V^e_{(\mu}u_{\nu)} 
	+ \tau_{\mu\nu} \\
	J_{\mu} &= \rhot u_{\mu} + \frac{\rhos}{\mu} P_{\mu}^{\phantom{\mu}\alpha} \partial_{\alpha}\phi + \Upsilon_{\mu} \\
	u^{\mu} \partial_{\mu} \phi &= -(\mu+\upsilon)\,
\end{align}
\end{subequations}
such that $V^e_{\mu} u^{\mu} = \Upsilon_{\mu} u^{\mu} = 0$ and $\tau_{\mu\nu}u^{\mu} = 0$.
The energy flux associated with $T_{\mu\nu}$ is
\begin{align}
\begin{split}
	j^e_{\mu} &= - P_{\mu\alpha}u_{\beta}T^{\alpha\beta} \\
		&= V^{e}_{\mu} + \rhos\left(\partial_{\mu}\phi - (\mu+\upsilon) u_{\mu} \right)
\end{split}
\end{align}
and the charge flux is
\begin{equation}
	j^c_{\mu} = \frac{\rhos}{\mu} \left(\partial_{\mu}\phi - (\mu+\upsilon)u_{\mu}\right) + \Upsilon_{\mu}\,.
\end{equation}
There is some freedom in choosing $\Upsilon_{\mu}$ and $V^e_{\mu}$.
We can specify the energy flux seen in the rest frame of the normal component (the Landau frame) or the charge density in the rest frame of the normal component (the Eckart frame). This amounts to choosing the direction of the velocity field relative to $V^e_{\mu}$ or $\Upsilon_{\mu}$. In what follows we define $u_{\mu}$ such that 
\begin{equation}
\label{E:DefLandau}
	V^e_{\mu}=0\,.
\end{equation} 
We will refer to this frame as the Landau frame. 
As in section \ref{S:Inviscid} we will use
\begin{equation}
	\mu_{\rm s} = -u^{\nu}\partial_{\nu}\phi 
\end{equation}
and
\begin{equation}
\label{E:Defnumu}
	\mu_{\rm s} n_{\mu} = P_{\mu}^{\nu}\partial_{\nu}\phi\,.
\end{equation}

The viscous corrections to the energy momentum tensor $\tau_{\mu\nu}$, charge current 
$\Upsilon_{\mu}$, and Josephson condition $\upsilon$
vanish when gradients of the hydrodynamic variables can be neglected. Thus, it is natural to expand $\tau_{\mu\nu}$, $\Upsilon_\mu$ and $\upsilon$ in gradients of $u^{\mu}$, $\partial^{\mu}\phi$, $\mu$ and $T$. In what follows we will consider these viscous corrections 
to linear order in gradients.  There are many distinct expressions which one can construct from the hydrodynamic variables. In principle all of these can contribute to $\tau_{\mu\nu}$, $\Upsilon_{\mu}$ and $\upsilon$. By distinct terms we mean ones that differ after the conservation equations $\partial_{\mu} T^{\mu\nu}=0$ and $\partial_{\mu} J^{\mu}=0$ have been implemented. 
In order to simplify our analysis of the viscous corrections we will only consider expressions which will play a role in a conformally invariant theory and only expressions for which $n^{\mu}$ is small, $n^\mu \ll 1$. The latter restriction will significantly reduce the number of possible terms that can be written down. It is a physically sensible restriction since superfluidity breaks down at large superfluid velocities. 

Within this approximation, the only possible Weyl covariant traceless symmetric tensors orthogonal to the velocity field that could contribute to $\tau_{\mu\nu}$ are
\begin{equation}
\label{E:sigmas}
	\sigma^{n}_{\mu\nu} = 2\partial_{\langle \mu}u_{\nu \rangle} \qquad
	\sigma^{s}_{\mu\nu}  = 2\partial_{\langle \mu}n_{\nu \rangle}\,.
\end{equation}
Here triangular brackets indicate a symmetric traceless projection onto the space orthogonal to the velocity field,
\begin{equation}
\label{E:SymmTraceless}
	A_{\langle \mu\nu\rangle} = \frac{1}{2} P_{\mu}^{\phantom{\mu}\lambda} P_{\nu}^{\phantom{\nu}\sigma} \left( A_{\lambda \sigma} + A_{\sigma \lambda} \right) - \frac{1}{3} P_{\mu\nu}P^{\lambda \sigma} A_{\lambda \sigma}\,,
\end{equation}
and $P_{\mu\nu}$ denotes a projection onto the space orthogonal to the normal velocity as in \eqref{E:Projectiondef}.
Assuming only parity even terms, the only possible Weyl covariant contribution to $\Upsilon^{\mu}$ is
\begin{equation}
	V_{\mu}^{n} = P_{\mu\nu}\partial^{\nu} \frac{\mu}{T} \,.
\end{equation}
The only Weyl covariant scalar contribution to $\upsilon$ is
\begin{equation}
\label{E:ViscousS}
	s^s = \partial_{\mu} (\rhos n^{\mu})\,.
\end{equation}
(While not directly obvious, both $u^{\mu} \partial_{\mu} \mu/T$ and $\partial_{\mu}( \rho_{\rm s} u^{\mu})$ are equivalent to $s^s$ once we require that $n^{\mu} \ll 1$ and that the energy momentum tensor and charge current are conserved.)
Within our approximation, the most general viscous corrections we can write are:
\begin{align}
\begin{split}
\label{E:ViscousCorrections}
	\tau_{\mu\nu} &= - \eta \sigma^{n}_{\mu\nu} - \eta_s \sigma^{s}_{\mu\nu} =
	- 2 \eta \partial_{\langle \mu}u_{\nu \rangle} - 2 \eta_s \partial_{\langle \mu}n_{\nu \rangle} \\
	\Upsilon_{\mu} & = -\kappa V_{\mu}^n  
	= -\kappa P_{\mu\nu}\partial^{\nu} \frac{\mu}{T}\\
	\upsilon & = -\zeta_3 s^{s} 
	= - \zeta_3 \partial_{\mu} (\rhos n^{\mu})\,.
\end{split}
\end{align}

One can attempt an alternate analysis of the possible viscous corrections by considering the combination 
\begin{equation}
	\mu \partial_{\mu}J^{\mu} + u_{\mu}\partial_{\nu} T^{\mu\nu} = 0
\end{equation}
and interpreting it as a statement regarding the entropy current similar to what was done in \eqref{E:SCurrent}--\eqref{E:InviscidParameters}. 
In the limit $n^{\mu} \ll 1$, it is possible to reorganize the terms of \eqref{E:ViscousHydro1} in a way that makes this entropy calculation simpler.  
We can redefine $\tau_{\mu\nu}$ and $\Upsilon_{\mu}$ so that in the Landau frame equation \eqref{E:ViscousHydro1} takes the form
\begin{subequations}
\label{E:ViscousHydro2}
\begin{align}
\label{E:InviscidT2}
	T_{\mu\nu} &= (\epsilon+P ) u_{\mu}u_{\nu} + P \eta_{\mu\nu} + 2 \rhos \mu_{\rm s} n_{(\mu} u_{\nu)} +\rhos \mu_{\rm s} n_{\mu}n_{\nu}  + \overline \tau_{\mu\nu} \\
\label{E:InviscidJ2}
	J_{\mu} &= \rhot u_{\mu} +\rho_{\rm s}n_{\mu} + \overline \Upsilon_{\mu} \\
	-\mu_{\rm s} \equiv u^{\mu} \partial_{\mu} \phi &= -(\mu+\upsilon)\,.
\end{align}
\end{subequations}
In the limit $n^\mu \ll 1$, $\Upsilon^\mu \approx \overline \Upsilon^\mu$
and $\tau^{\mu\nu} \approx \overline \tau^{\mu\nu}$.  

The authors of \cite{Pujol:2002na,Gusakov:2007px}, 
found that the form \eqref{E:ViscousHydro2} together with the relations 
\begin{equation}
\label{E:newdP}
	dP = sdT + \rhot d\mu -\frac{\rhos}{2 \mu_{\rm s}} d(\mu_{\rm s}^2 n^2)
\end{equation}
and
\begin{equation}
\label{E:phiSymmetries}
	\partial_{\nu} (\mu_{\rm s} n_{\mu}) - \partial_{\mu} (\mu_{\rm s} n_{\nu})
	=
	-\partial_{\nu} (\mu_{\rm s} u_{\mu}) + \partial_{\mu} (\mu_{\rm s} u_{\nu})
\end{equation}
(which follows from \eqref{E:Defnumu}) 
implies
\begin{align}
\begin{split}
\label{E:viscousudT}
	\mu \partial_{\nu} J^{\nu} + u_{\mu} \partial_{\nu} T^{\mu\nu} 
	& = -T \partial_{\nu} \left( s u^{\nu} - \frac{\mu}{T} \Upsilon^{\nu} \right)  \\
	&-\upsilon  \partial_{\nu} \left( \rhos n^{\nu} \right) - T \Upsilon^{\nu} \partial_{\nu} \frac{\mu}{T} - \tau^{\mu\nu}\partial_{\nu}u_{\mu}\,.
\end{split}
\end{align}
Following Landau \cite{LandL}, if we interpret the term in the parentheses on the first line of \eqref{E:viscousudT} as the entropy current $J^{\mu}_s$ and relax the assumption of conformal invariance then we find that the second law, $\partial_{\mu} J_s^{\mu} \geq 0$, implies 
\begin{align}
\begin{split}
	\Upsilon_{\mu} &= - \kappa V_{\mu}^n \\
	\upsilon &= -\zeta_3 \partial_{\sigma} \left(\rho_{\rm s} n^{\sigma} \right) - \zeta_2 \partial_{\sigma}u^{\sigma} \\
	\tau_{\mu\nu} & = -\eta \sigma^{n}_{\mu\nu} - \zeta_1 P_{\mu\nu} \partial_{\sigma}u^{\sigma} - \zeta_2 P_{\mu\nu} \partial_{\sigma} \left(\rho_{\rm s} n^{\sigma} \right)
\end{split}
\end{align}
where $\eta>0$, $\kappa>0$ and either $\zeta_1 >0$, $\zeta_2 \leq \zeta_1 \zeta_3$ or $\zeta_1 = \zeta_2 = 0$ and $\zeta_3>0$. Imposing conformal invariance implies that we must choose the latter which coincides with \eqref{E:ViscousCorrections} with $\eta_{\rm s}=0$. The observation that $\zeta_3$ does not necessarily vanish in a conformal theory was made in the non-relativistic case in \cite{Son:2005tj}. Before ending this section we note that one should be mindful of the ad hoc interpretation of the expression in the parentheses on the first line of \eqref{E:viscousudT} as the entropy current. In \cite{Erdmenger:2008rm,Banerjee:2008th,Son:2009tf} it was shown how such an interpretation fails in parity violating theories. Also, \eqref{E:newdP} is somewhat non standard since it implies that the dependence of the thermodynamic quantities like pressure and energy on $\mu$, $T$ and $(\partial\phi)^2$ gets corrected by viscous effects.

To summarize, we expect that for $n_{\mu} \ll 1$
\begin{subequations}
\label{E:fullequations}
\begin{align}
\label{Tmunufull}
	T_{\mu\nu} &= (\epsilon+P)u_{\mu}u_{\nu} + P  \eta_{\mu\nu} + 2\rhos \mu_{\rm s} n_{(\mu}u_{\nu)} + \rhos \mu_{\rm s} n_{\mu}n_{\nu} - \eta \sigma_{\mu\nu}^n -\eta_s\sigma_{\mu\nu}^s \\
\label{Jmufull}
	J_{\mu} & = \rhot u_{\mu} + \rhos n_{\mu} -\kappa V^n_{\mu} \\
\label{E:Josephson}
	-\mu_{\rm s} \equiv u^{\mu} \partial_{\mu}\phi &= -\mu+\zeta_3 s^s
\end{align}
\end{subequations}
where $\mu_{\rm s} n_{\mu} = P_{\mu}^{\nu} \partial_{\nu}\phi$.
In the constitutive relations \eqref{E:fullequations},
we have kept the full dependence on $n^\mu$ at zeroth order in the gradient expansion.  At first order in the gradient expansion, we have discarded any terms which contain an $n^\mu$ 
that is not acted on by a derivative.

\subsection{Kubo Formulas}
\label{S:Kubo}
In this section we deduce some Kubo relations for the retarded Green's functions,
\begin{align}
\begin{split}
G^{\mu\nu, \lambda \sigma}_R(\omega, k) &= i \int d^4x \, e^{ - i k x} \theta(t) \left\langle [ T^{\mu\nu}(x), T^{\lambda \sigma}(0) ] \right\rangle \\
G^{\mu,\nu}_R(\omega, k) &= i \int d^4x \, e^{- i k x} \theta(t) \left\langle [ J^\mu(x), J^\nu(0) ] \right\rangle \\
G^{\phi\phi}_R(\omega, k) &= i \int d^4x \, e^{ - i k x} \theta(t) \left\langle [ \phi(x), \phi(0) ]\right\rangle \,,
\end{split}
\end{align}
from linearized hydrodynamics.
A similar analysis with somewhat different notation and different methods can be found in \cite{Valle:2007xx}.

Let us look at small fluctuations about an equilibrium state at fixed temperature, chemical potential and zero normal and superfluid velocities, i.e.\ we write 
\begin{equation}
\label{E:perturbations}
	T = T_0 + T' \qquad 
	\mu = \mu_0 + \mu' \qquad
	u^\mu = (1, u^i) \qquad
	n^\mu = (0, (\partial^i \phi)/\mu - u^i)\,
\end{equation}
where $T'$, $\mu'$, $u_i$ and $\partial_i\phi$ are small.
At linear order in the fluctuations, the stress tensor (\ref{Tmunufull}), current (\ref{Jmufull}), and phase (\ref{E:Josephson}) have the form
\begin{subequations}
\begin{align}
T_{00} &= \epsilon \\
T_{0i} &= -(\mu \rhon + sT) u_i - \rhos \partial_i \phi  \\
T_{ij} &= (P + \frac{ 2\eta}{3} \partial_k u^k ) \delta_{ij} -  2\eta \partial_{(i} u_{j)} 
\label{Tijlinear}
\\
J_0 &= -\rhot \\
J_i &= u_i \rhon + \frac{\rhos}{\mu} \partial_i \phi - \frac{\kappa}{T} \left( \partial_i \mu - \frac{\mu}{T} \partial_i T \right)  
\label{Jilinear}
\\
\partial_0 \phi &= - \mu + \zeta_3 \rhos \left( \frac{\partial_i \partial^i \phi}{\mu}  - \partial_i u^i\right) \ .
\label{philinear}
\end{align}
\end{subequations}

From (\ref{Tijlinear}), we can deduce the standard Kubo relation for the viscosity $\eta$: a velocity $u_x(y)$ can be thought of as a small Galilean boost $x \to x - u_x(y) t$, which in turn leads to a metric fluctuation $\delta g_{xy}$ such that $\partial_0 \delta g_{xy} = \partial_y u_x$.  Thus
\begin{equation}
\lim_{\omega \to 0} \frac{1}{\omega} \operatorname{Im} G^{xy,xy}(\omega,0) = \eta \,.
\end{equation}

A Kubo formula for $\kappa$ can be obtained by the identification $\partial_i \mu = -\partial_i A^{\hbox{\tiny ext}}_t$ in (\ref{Jilinear}) where $A^{\hbox{\tiny ext}}$ is an external gauge field.  The long wave-length limit of the current-current correlation function is then:
\begin{equation}
\lim_{\omega \to 0} \lim_{k \to 0} \frac{1}{k} \operatorname{Im} G^{x,t}(\omega, k) = \frac{\kappa}{T} \ .
\label{KubokappaT}
\end{equation}
One more Kubo relation, also discussed in \cite{Herzog:2009md}, can be extracted from (\ref{Jilinear}) by replacing $\partial_i\phi$ with the gauge invariant combination $\partial_i \phi - A^{\hbox{\tiny ext}}_i$:
\begin{equation}
\lim_{\omega \to 0}  \operatorname{Re} G^{x,x}(\omega, 0) = -\frac{\rhos}{\mu} 
\ .
\label{Kuborhosmu}
\end{equation}
Finally, we derive a Kubo formula for $\zeta_3$.  Using current conservation, we write (\ref{philinear}) in the more suggestive form
\begin{equation}
\label{E:PerturbedJosephson}
\partial_0 \phi = - \mu_0 - \frac{\partial\mu}{\partial \rhot} J^0 - \zeta_3 \partial_0 J^0 + \ldots 
\end{equation}
where by $\ldots$ we mean second order terms in a gradient expansion.
If we turn on an external gauge field $A^{\hbox{\tiny ext}}_0$ we expect that the left hand side of \eqref{E:PerturbedJosephson} will be modified to $\partial_0 \phi - A^{\hbox{\tiny ext}}_0$. We can think of the variable $J^0$ as being canonically conjugate to $-\partial_0 \phi$.  This conjugacy allows us to deduce the relations
\begin{eqnarray}
\lim_{\omega \to 0} \frac{1}{\omega^2} \operatorname{Re} G^{\phi \phi} (\omega,0) &=& -\frac{\partial \mu}{\partial \rho} 
\label{Kubomurho}
\\
\lim_{\omega \to 0} \frac{1}{\omega} \operatorname{Im} G^{\phi \phi} (\omega,0) &=& \zeta_3 \ .
\label{Kubozeta3}
\end{eqnarray}

\subsection{Sound attenuation}
\label{S:Sound}
Superfluids admit several types of sound modes \cite{LandL,Atkins}. In addition to carrying sound by pressure waves, as normal fluids do, superfluids allow for a second sound mode through entropy waves.    A third sound mode can be generated by surface waves on a thin film of superfluid and a fourth sound mode exists when the normal component of the velocity field is prevented from moving. First, second and fourth sound were studied in a relativistic setting in the context of AdS/CFT in \cite{Herzog:2008he,Yarom:2009uq,Herzog:2009md}. In what follows we will study the attenuation of first, second and fourth sound due to $\kappa$, $\zeta_3$ and $\eta$.

At linearized order in the fluctuations described in \eqref{E:perturbations}, the conservation of the stress tensor, $\partial_\mu T^{\mu 0}=0$ and $\partial_\mu T^{\mu i}=0$,  and of the current, $\partial_\mu J^\mu = 0$, imply that
\begin{subequations}
\begin{align}
 \left( \mu \frac{\partial\rhot}{\partial \mu} + T \frac{\partial s}{\partial \mu} \right) \partial_0 \mu'
+ \left( \mu \frac{\partial\rhot}{\partial T} + T \frac{\partial s}{\partial T} \right) \partial_0 T'
+ (\mu \rhon + s T) \partial_i u_i + \rhos \partial_i^2 \phi &= 0 
\label{dmuTmu0}
\\
\rhot \partial^i \mu' + s \partial^i T'+
(\mu \rhon + s T) \partial_0 u^i + \rhos \partial_0 \partial^i \phi 
- \frac{ \eta}{3} \partial^i \partial_j u^j - \eta \partial_j^2 u^i &= 0 
\label{dmuTmui}
 \\
 \frac{\partial \rhot}{\partial \mu} \partial_0 \mu' + \frac{\partial \rhot}{\partial T} \partial_0 T' + 
\rhon \partial_i u^i + \frac{\rhos}{\mu} \partial_i^2 \phi - \frac{\kappa}{T}
\left( \partial_i^2 \mu' - \frac{\mu}{T} \partial_i^2 T' \right) &= 0 
\label{dmuJmu}
\ .
\intertext{Another important relation in constructing the dispersion relation for the sound modes is the derivative $\partial_i$ of (\ref{philinear}):}
\partial_0 \partial_i \phi + \partial_i \mu' - \zeta_3 \rhos
\left( \frac{\partial_i \partial_j^2 \phi}{\mu} - \partial_i \partial_j u^j \right) &=0 \ .
\label{did0phi}
\end{align}
\end{subequations}
Assuming the space-time dependence of the fluctuations $X^T = (\mu', T', u^x, \partial^x \phi)$ takes the form 
$e^{-i \omega t + i k x}$, we can write (\ref{dmuTmu0}), (\ref{dmuTmui}), (\ref{dmuJmu}), and (\ref{did0phi}) schematically as a linear system:
\begin{equation}
	M X = 0 \ ,
\end{equation}
where $M$ is a 4$\times$4 matrix that depends on $\omega$ and $k$.

As long as the momentum $k$ is small enough, the dispersion relations for first and second sound can be determined from the four roots
\begin{equation}
\label{E:roots}
\omega = \pm c_a k - i k^2 \Gamma_a + \ldots,
\end{equation}
$a=1,2$ 
of the determinant of $M$. To find the roots, 
we introduce the entropy per particle $\sigma = s / \rho$ and  take advantage of conformal invariance.  
Using the scaling form for the pressure, $P = T^4 f(\mu/T)$, we express all the susceptibilities $\partial \rhot / \partial T$, $\partial \rhot / \partial \mu$, $\partial s / \partial T$, and $\partial s / \partial \mu$ in terms of $\partial \sigma / \partial T$.  For instance
\begin{equation}
\frac{\partial \rhot}{\partial T} = \frac{\partial s}{\partial \mu} =
\frac{\rho}{\mu \rhot + s T} \left( 3s - \rhot T \frac{\partial \sigma}{\partial T} \right) \ .
\end{equation}
At leading order in $\omega$ and $k$, the determinant of $M$ has the form
\begin{equation}
\det ( M ) =  (k^2 - 3 \omega^2) \left(k^2 \rhos \sigma^2 - (\mu \rhon + s T) \frac{\partial \sigma}{\partial T}\omega^2 \right) + \ldots \ ,
\end{equation}
from which we can read off the sound speeds
\begin{equation}
	c_1^2 = \frac{1}{3} \ ,
	\qquad
	c_2^2 = \frac{\rhos \sigma^2}{\left(\mu \rhon + s T \right) \frac{\partial \sigma}{\partial T}}\,.
	\label{c1andc2}
\end{equation}
See \cite{Herzog:2009md} for details. 

The viscous corrections discussed in section \ref{S:Viscous} allow us to determine the sound attenuation $\Gamma_a$, $a=1,2$.
By considering the first subleading corrections to $\det(M)$ we find that
\begin{equation}
	\Gamma_1 =  \frac{2 \eta}{3 (\mu \rhot + s T)} 
\end{equation}
for first sound, which is the standard expression for sound attenuation in a conformally invariant theory. For second sound we find 
\begin{equation}
	\Gamma_2 =  \frac{2  \mu \rhos}{3 (\mu \rhot + s T) (\mu \rhon + s T)}\eta
	+ \frac{ \mu (\mu \rhot + s T)}{2 (\partial \sigma / \partial T) \rhot^2 T^3} \kappa
	+ \frac{ \rhos (\mu \rhot + s T)}{2 \mu (\mu \rhon + s T)} \zeta_3 \ 
	.
	\label{Gamma2}
\end{equation}

The dispersion relation \eqref{E:roots} is valid only as long as the momenta are small $\Gamma_a k \ll c_a$. 
When $\rho_{\rm s}=0$, the speed of second sound vanishes, and
 the roots \eqref{E:roots} reduce to
\begin{equation}
\label{E:Tcroots}
	\omega = \pm c_1 -i k^2 \Gamma_1 \qquad
	\omega = -i k^2  \frac{ \mu (\mu \rhot + s T)}{(\partial \sigma / \partial T) \rhot^2 T^3} \kappa\,,
\end{equation}
provided $\Gamma_1 k \ll c_1$.  (In this limit, the fourth root sits at $\omega=0$.)

Fourth sound is associated with wave propagation of the superfluid component when the normal component is forced to stay motionless. Experimentally, such a configuration is realized by channeling the superfluid through a tube packed with fine powder which immobilizes the normal component. In this setup the energy momentum tensor of the superfluid is not conserved since momentum is transferred to the medium which holds the normal component in place. Thus, we consider linear perturbations of the form $X = (\mu', \partial^i \phi)$, keeping $T$ and $u^\mu = (1,0)$ fixed. 
In this limit, we only need to consider (\ref{dmuJmu}) and (\ref{did0phi}) which reduce to
\begin{align}
\begin{split}
\label{E:sound4}
\frac{\partial \rhot}{\partial \mu} \partial_0 \mu' + \frac{\rhos}{\mu} \partial_i^2 \phi - \frac{\kappa}{T} \partial_i^2 \mu' &= 0 \\
\partial_0 \partial_i \phi + \partial_i \mu' -
\zeta_3 \frac{\rhos}{\mu} \partial_i \partial_j^2 \phi
&= 0 \ .
\end{split}
\end{align}
We assume $\mu'$ and $\partial_i \phi$ both have a $e^{-i \omega t + i k x}$ dependence.  The determinant of the system of equations \eqref{E:sound4} is the polynomial
\begin{equation}
	\det(M) = \omega^2  - c_4^2 k^2 +2 i k^2 \Gamma_4 \omega + \ldots
\end{equation}
where
\begin{equation}
c_4^2 =\frac{\rhos}{\mu (\partial \rhot / \partial \mu)} \ , \; \; \;
\Gamma_4 =  
\frac{1}{2 T (\partial \rhot/\partial\mu)} \kappa + \frac{\rho_s}{2 \mu} \zeta_3 \ 
\end{equation}
and $\ldots$ denote subleading terms in the momenta.

Expanding out the roots for small momenta, $\Gamma_4 k \ll c_4$, one finds
\begin{equation}
\omega = \pm c_4 k - i \Gamma_4 k^2 + \mathcal{O}(k^3) \ .
\label{fourthsounddispersion}
\end{equation}
On the other hand, when $\rhos=0$ we have
\begin{equation}
\label{E:fourthsoundAlternate}
	\omega = - i k^2 \frac{\kappa}{T (\partial \rhot/\partial\mu)}\,.
\end{equation} 
As expected, these dispersion relations agree with those associated with second sound in the limit where $sT \gg \mu \rhot$.

\section{The method of computation}
\label{S:method}
Holographic superconductors were initially constructed in  \cite{Hartnoll:2008vx} following the pioneering work of \cite{Gubser:2008px}. The simplest construction of a superfluid involves a charged scalar, a $U(1)$ field and a graviton, with an action
\begin{equation}
\label{E:action}
	S = \frac{1}{2\kappa_5^2} \int \sqrt{-g} \left(R + \frac{12}{L^2} -\frac{1}{4}F^{mn}F_{mn} - |\partial_{m}\psi - i q A_{m} \psi|^2 - V (|\psi|^2) \right)d^5x + S_b\,.
\end{equation}
Roman indices $m,\,n=0,\ldots 3,5$ are raised and lowered with the bulk metric. 
The index 5 refers to the radial coordinate.\footnote{%
 We also use the boundary indices $\mu$, $\nu = 0, \ldots, 3$ and $i, j = 1, 2, 3$ from before.
 The indices $\mu$, $\nu$  and $i$, $j$ are  
 raised and lowered with the Minkowski metric $\eta_{\mu\nu}$ and with $\delta_{ij}$ respectively.
 }  
In \eqref{E:action} $g_{mn}$ is the metric, $R$ is the Ricci scalar, $F = dA$ is the field strength of the $U(1)$ gauge field,
and $\psi$ is a complex scalar with charge $q$. The AdS radius is denoted $L$ and we will set it to $1$ in what follows. $S_b$ denotes boundary terms which do not affect the equations of motion but will affect gauge theory correlators. At low enough temperatures and large enough charge, solutions to the equations of motion following from \eqref{E:action} admit a condensed configuration where $\psi$ is non vanishing. This is the holographic dual of the superfluid phase. Some introductory material to holographic superconductors can be found in \cite{Hartnoll:2009sz,Herzog:2009xv,Horowitz:2010gk,Sachdev:2010ch}. 

The goal of the remainder of this paper is twofold: To construct a holographic dual for a space-time dependent superfluid, similar in spirit to the construction of \cite{Bhattacharyya:2008jc}, and also to compute the transport coefficients $\eta$, $\eta_s$, $\kappa$ and $\zeta_3$ associated with it. In what follows we will summarize our method of computation.

In the simplest setup, one considers a superfluid in which the normal component and the superfluid component are motionless. The solution to the equations of motion in this case involve the metric $g_{\mu\nu}$, $A_0$, and $\psi$ (and perhaps $A_5$, depending on our gauge choice). We find it useful to work with gauge invariant variables $G_m = A_m -  \partial_m \varphi $ where $\varphi$ is the phase of $\psi$,
\begin{equation}
	\psi = \frac{1}{\sqrt{2}} \rho e^{i q \varphi}\,.
	\label{E:psirhophi}
\end{equation}
The AdS/CFT correspondence relates $g_{\mu\nu}$ to the metric and energy momentum tensor of the boundary theory and $G_0$ to the chemical potential and charge density.
A particularly useful relation that we derive is that, in our conventions, the boundary value of $G_0$ gives us the Josephson condition \eqref{E:Josephson}. All these relations will be made more precise in sections \ref{S:Static} and \ref{S:Dynamical} and in appendix \ref{A:BtoB}.

To obtain dynamics, we need that the boundary theory include non-zero superfluid and normal component velocities.
At first we will be interested in an inviscid superfluid, i.e.\ one where gradient corrections can be neglected or are non-existent. Viscous effects are absent if both the superfluid and normal component velocities are constant in space and time.
By turning on a constant boundary value of $G_i$, we introduce a non vanishing but constant superfluid velocity \cite{Herzog:2008he}.
In the condensed phase, a non zero superfluid velocity implies a non vanishing value for the spatial component of the charged current. 
Following our analysis of superfluids in section \ref{S:Superfluids} we will only be interested in configurations with a small superfluid velocity. Thus we consider only linear perturbations of $G_i$ around the static solution described earlier.

To obtain a solution where both the superfluid component and the normal component are in motion, we Lorentz boost the boundary theory energy momentum tensor and current. After such a boost the normal component and the superfluid component are in motion. Such a boost can be achieved in the gravitational dual by a coordinate transformation.

So far we have a gravitational description of a superfluid with arbitrary constant superfluid velocity and arbitrary but constant velocity for the normal component. These configurations have been studied in the literature in \cite{Herzog:2008he,Basu:2008st,Yarom:2009uq,Herzog:2009md,Amado:2009ts,Gubser:2009qf,Keranen:2009ss,Keranen:2009re,Sonner:2010yx,Arean:2010wu,Keranen:2010sx}. To go beyond these stationary solutions, we need to allow for the velocity fields and other thermodynamic quantities to be spacetime dependent. When we discussed viscous corrections in the boundary theory in section \ref{S:Viscous}, we constructed these corrections to first order in gradients of the thermodynamic variables. It is natural to carry out the same kind of analysis in the gravitational dual. The stationary configuration is described by the metric $g_{mn}$, the gauge invariant combination $G_m$ and the modulus of the scalar field $\rho$. These fields encode in them the boundary theory velocity fields $n_{\mu}$ and $u_{\mu}$, the chemical potential $\mu$ and the boundary temperature $T$ which are all constants independent of the boundary coordinates $x^{\mu}$. If we promote all these parameters to be spacetime dependent $n_{\mu} = n_{\mu}(x^{\nu})$, $u_{\mu} = u_{\mu}(x^{\nu})$, etc., then the modified fields $g_{mn}$, $A_m$ and $\rho$ will no longer solve the Einstein equations and matter equations. To correct this, we add corrections to the metric $\delta g_{mn}$, gauge field $\delta G_m$, and scalar $\delta \rho$ such that the combinations $g_{mn} + \delta g_{mn}$, $G_m + \delta G_m$ and $\rho+\delta\rho$ solve the equations of motion. It is difficult to find the delta'd quantities in full generality, but if we focus only on first order gradient corrections to the superfluid, then it is sufficient to find $\delta g_{mn}$, $\delta G_m$ and $\delta \rho$ which will solve the equations of motion to linear order in the gradients of $n_{\mu}$, $u_{\mu}$, $T$ and $\mu$.

Carrying out this computation we achieve our first goal, to construct a holographic dynamical superfluid. To compute the various transport coefficients, we use the AdS/CFT dictionary, described in appendix \ref{A:BtoB}. The energy momentum tensor of the superfluid is dual to $g_{mn} + \delta g_{mn}$. From it, we can read off the shear viscosities $\eta$ and $\eta_s$. The charged current is dual to $G_m$ and from it we can compute the diffusion coefficient $\kappa$. Finally, we obtain $\zeta_3$ by evaluating the boundary value of $u^{\mu} \delta G_{\mu}$.

The computation described above is feasible in principle but in practice technically difficult. While we are able to make some general statements about the holographic superfluid (e.g.\ $\eta_s=0$ and $\zeta_3$ diverges close to the phase transition), it is difficult to obtain explicit expressions for $\zeta_3$ and the other transport coefficients. 
The main obstacle is the absence of analytic control over the static solution. To overcome this problem we use the recently discussed analytic holographic superfluid of \cite{Herzog:2010vz}. In \cite{Herzog:2010vz} an analytic solution to the equations of motion were obtained in the limit where the metric does not backreact on the matter fields and for temperatures close to the 
phase transition. In other words, the solution of \cite{Herzog:2010vz} involves a double expansion. One parameter---the charge of the scalar field---allows for the matter fields to be weak so that they do not interact strongly with gravity.\footnote{%
 This same ``probe limit" was used in one of the first holographic superconductor papers \cite{Hartnoll:2008vx}.
} 
The other parameter is the distance from the phase transition, $T/\mu - (T/\mu)_0$, with $(T/\mu)_0$ the critical value of $T/\mu$ at which the phase transition occurs.\footnote{%
 In \cite{Herzog:2010vz}, 
  the magnitude of the order  parameter replaced $T/\mu - (T/\mu)_0$ 
  as the small parameter of the expansion.
 }
Thus, in our formulation there is a triple expansion involving the charge of the scalar, $q$, the relative  temperature, $T/\mu-(T/\mu)_0$, and gradients, $\delta$. 

First, neglecting gradient corrections, in section \ref{S:Static} we describe the static solution. 
To leading order in the charge of the scalar, the metric completely decouples from the matter fields and the solution to the Einstein equations is the Schwarzschild black hole which we describe in \ref{S:AdSSS}. At the next order in $q$, we solve for the matter fields. This is described in section \ref{S:Probe}.
We then go beyond the probe limit in section \ref{S:backreaction} where we consider the leading order backreaction of the metric. Up to section \ref{S:Boosted}, our analysis is general. We express the particular analytic solution of \cite{Herzog:2010vz} as an expansion in $T/\mu-(T/\mu)_0$ in section \ref{S:Explicit}. Section \ref{S:Dynamical} extends the solution of section \ref{S:Static} to include gradient corrections \`a la \cite{Bhattacharyya:2008jc}. The extension is performed order by order in $q$ in sections \ref{S:GEPO0} to \ref{S:GEPO2}.  The principal results are that $\zeta_3$ diverges at $(T/\mu)_0$, $\kappa$ is continuous across the phase transition, $\eta/s=1/4\pi$, and $\eta_s=0$. For the analytic solution of \cite{Herzog:2010vz}, the explicit values of $\zeta_3$ and $\kappa$ described in \eqref{E:FinalKappa} and \eqref{E:FinalZeta3} are obtained in a perturbative expansion near $T/\mu-(T/\mu)_0$ in \ref{S:ExplicitO1}.

\section{Static holographic superfluids}
\label{S:Static}

Before describing the solution to the equations of motion that follow from (\ref{E:action}), 
we need to explain the precise form of the boundary action $S_b$ and also the relationship
between the bulk gravity fields $g_{\mu\nu}$, $A_\mu$ and $\psi$ and the boundary field theory quantities $T^{\mu\nu}$, $J^\mu$, and $\phi$.  

The boundary action $S_b$ can be obtained by requiring that the variational principle is well defined. 
Assuming that the metric $g_{mn}dx^m dx^n$ becomes asymptotically anti-de Sitter at large $r$,
\begin{equation}
\label{E:AdSAsymptotics}
	\lim_{r \to \infty} ds^2 = -r^2 dt^2 + r^2 \sum_i (dx^i)^2 + 2 dr dt \,,
\end{equation}
we take $S_b$ to live on a constant and large $r$ surface.  On this surface, we can define an induced metric $\gamma_{mn}$ and a unit vector $n^m$ (the lapse function, not to be confused with the relative superfluid velocity $n^{\mu}$) that points toward larger $r$.
The terms in $S_b$ relevant for this paper are
\begin{align}
\begin{split}
\label{E:Sboundary}
	S_b =& \frac{1}{2\kappa_5^2}\int_{\tiny \hbox{boundary}} \sqrt{-\gamma} \left(2 K - 6\right) d^4x 
	\\
		&+\frac{1}{2\kappa_5^2} \int_{\tiny \hbox{boundary}} \sqrt{-\gamma} \left( m_{\Delta}^2 |\psi|^2 +\frac{1}{2} \tilde{m}_{\Delta}^2 \left(\psi^* n^{m}\partial_{m} \psi + \hbox{c.c} \right) \right) d^4 x \ .
\end{split}
\end{align}
Here $K_{\mu\nu}$ is the extrinsic curvature,
$
	K_{\mu\nu} = - \Gamma^5_{\mu\nu} / \sqrt{g^{55}}
$
with $\Gamma^{m}_{np}$ the Christoffel symbol, and $K = K_{\mu\nu} \gamma^{\mu\nu}$.

The expressions for $m_{\Delta}$ and $\tilde{m}_{\Delta}$ depend on the scalar potential 
\begin{equation}
\label{E:scalarpotential}
	V(|\psi|^2)  = m^2 |\psi|^2  + \mathcal{O}(|\psi|^4)\,,
\end{equation}
in (\ref{E:action}).  Through the AdS/CFT dictionary, the mass of the scalar $\psi$ is related to the conformal dimension $\Delta$  of the dual operator $O_{\psi}$ through
\begin{equation}
	m^2 = \Delta (4-\Delta)\,.
\end{equation}
By requiring a well posed variational problem and also that the boundary action is finite, one finds that $m_{\Delta}$ and $\tilde{m}_{\Delta}$ in \eqref{E:Sboundary} must satisfy 
\begin{equation}
\label{E:mmt}
	\tilde{m}_{\Delta}^2 = \begin{cases}	0 & \Delta>2 \\ 2 & \Delta \leq 2\end{cases}
	\qquad
	\hbox{and}
	\qquad
	m_{\Delta}^2 = \begin{cases} \Delta-4 & \Delta>2 \\ \Delta & \Delta \leq 2 \end{cases}.
\end{equation}

The prescription for obtaining the energy momentum tensor in the dual (boundary) gauge theory, $T_{\mu\nu}$, was derived in \cite{Balasubramanian:1999re,Bianchi:2001de} following the earlier work of \cite{Gubser:1998bc,Witten:1998qj}.
For configurations in which the fields depend only on the radial coordinate $r$,
\begin{multline}
\label{E:gmnToTmn}
	\kappa^2_5 \langle T_{\mu\nu} \rangle  =  -\lim_{r \to \infty} r^2 \Bigg( K_{\mu\nu} - K \gamma_{\mu\nu} + 3 \gamma_{\mu\nu} 
	- \frac{1}{2} \gamma_{\mu\nu} m_{\Delta} |\psi|^2 
\\
	- \frac{1}{4} \gamma_{\mu\nu} \tilde{m}_{\Delta} \left(\psi^* n^{\mu}\partial_{\mu} \psi  - 2\psi^*n_{(\mu}\partial_{\nu)} \psi 
	+ \hbox{c.c.} \right)
\Bigg) \,.
\end{multline}

To understand the relationship between $O_{\psi}$ and $\psi$, note that
$\psi$ has the large $r$ expansion
\begin{equation}
\label{E:psiexpansion}
	\psi = \psi_s r^{-(4-\Delta)} \left(1 + \ldots\right)  + \psi_r {r^{-\Delta}} \left(1 + \ldots\right)\,,
\end{equation}
with $\ldots$ denoting subleading powers of $r$.\footnote{%
  For certain values of $\Delta$ the series expansion associated with $\psi_s$ and the series
   expansion associated with $\psi_r$ overlap. In this case one obtains logarithmic terms in the
   series expansion. We will see an explicit example of this sort of behavior in section \ref{S:Explicit}.
} 
The coefficient $\psi_s$ is associated with a source term for the operator $O_{\psi}$ and the coefficient $\psi_r$ is associated with $\langle O_{\psi} \rangle$. If we set $\psi_s=0$ implying that we are not deforming the boundary theory, and we consider configurations which depend only on the radial coordinate $r$, then the expectation value of the operator $O_{\psi}$ dual to $\psi$ is given by
\begin{equation}
\label{E:psiToO}
	\kappa_5^2 O_{\psi} =  \sqrt{2}C_{\Delta} \psi_r\,,
\end{equation}
where $C_{\Delta}$ denotes a dimensionless constant whose value depends on our conventions for normalizing the source term $\psi_s$ \cite{Klebanov:1999tb}. Its explicit value will not play an important role in this work.  We split $\psi$ into its modulus $\rho$ and phase $\varphi$ as in
(\ref{E:psirhophi}). 
Following \eqref{E:psiexpansion} and \eqref{E:psiToO}, we identify  the goldstone boson $\phi$ with  the phase of $O_{\psi}$  (in the absence of sources) through
\begin{equation}
\label{E:goldstone}
	 \langle \phi \rangle = \lim_{r \to \infty} \varphi \,,
\end{equation}
and the modulus of $O_{\psi}$ (in the absence of sources) with the near boundary asymptotics of $\rho$,
\begin{equation}
\label{E:GotOpsi}
	\kappa_5^2 |\langle O_{\psi} \rangle|  = C_{\Delta} \lim_{r \to \infty} r^{\Delta} \rho  \,.
\end{equation}

The boundary theory current $J_{\mu}$ is related to the bulk gauge field $A_{\mu}$ through
\begin{equation}
\label{E:AToJ}
	  \kappa_5^2 \langle J_{\mu} \rangle  = \lim_{r \to \infty} r^2 A_{\mu} 
	+ \pdt
\end{equation}
which is valid for gauge fields which depend only on the radial coordinate $r$ and the  gauge choice $A_5=0$.
If the configuration admits an event horizon at some $r b=1$ then the chemical potential of the boundary theory can be obtained from 
\begin{equation}
\label{E:Gotmu}
	\mu = \int_{1/b}^{\infty} \partial_r A_{t} dr \,.
\end{equation}

The results \eqref{E:gmnToTmn}, \eqref{E:psiToO} and \eqref{E:AToJ} can be modified to include more intricate spacetime dependent configurations by including appropriate boundary counterterms. In general, we expect these counterterms to depend on two or more derivatives of the bulk fields in the directions transverse to $r$. For the analysis carried out in this work, all such terms can be neglected and we will use \eqref{E:gmnToTmn}, \eqref{E:psiToO} and \eqref{E:AToJ} from now on.

Bulk configurations with $\psi=0$ and $A_0 \neq 0$ correspond to boundary theory configurations with a non vanishing charge density. Once $\psi$ is non vanishing and regular, and $\psi_s =0$ then the boundary configuration corresponds to a spontaneously broken phase of the theory where $\phi$, the phase of $O_{\psi}$, is the goldstone boson. 

To simplify our computations we formally expand the metric and matter fields in inverse powers of the scalar charge and work in the large charge ($q\to\infty$) limit:
\begin{equation}
\label{E:qansatz}
	g_{mn} = g_{mn}^{(0)}  +g_{mn}^{(2)} q^{-2} +  \mathcal{O}(q^{-4}) \qquad
	A_{m} = A_{m}^{(1)} q^{-1} + \mathcal{O}(q^{-3}) \qquad
	\psi = \psi^{(1)} q^{-1} + \mathcal{O}(q^{-3})\,.
\end{equation}
The ansatz \eqref{E:qansatz} is useful since the bulk energy momentum tensor of the matter fields  is quadratic in $\psi$ and $A_{\mu}$, effectively decoupling the Einstein equations from matter at leading order. In sections \ref{S:AdSSS}, \ref{S:Probe}  and \ref{S:backreaction} we will discuss the construction of a superfluid order by order in the large charge expansion. In section \ref{S:Boosted} we extend the solution to include a non-trivial velocity for the normal component. In section \ref{S:Explicit} we show how to construct an explicit solution to the field equations to order $q^{-2}$ close to the critical temperature.
We emphasize that in this section all viscous corrections to the stress tensor and current vanish because
the solutions presented  correspond to a fluid moving with uniform superfluid velocity, normal velocity, temperature and chemical potential.

\subsection{The AdS-Schwarzschild black hole.}
\label{S:AdSSS}
At order $q^0$ the solution to the Einstein equations is the AdS-Schwarzschild black hole given by the line element
\begin{equation}
\label{E:MetricAnsatz}
	ds^2 = r^2 \left( - \left(1 - \frac{1}{r^4 b^4}\right) dt^2 + \sum_{i=1}^3 (dx^i)^2\right)  + 2 dr dt\,.
\end{equation}
The black hole horizon is located at $r=1/b$ and the asymptotically AdS boundary is located at $r\to \infty$. The Hawking temperature of the black hole, which is also the temperature of the boundary theory is given by
\begin{equation}
\label{E:HawkingT}
	T  = \frac{1}{\pi b}\,.
\end{equation}
We have chosen to write the metric in ingoing Eddington-Finkelstein coordinates in order to avoid a coordinate singularity at the event horizon. 

Using the prescription \eqref{E:gmnToTmn} we find that the configuration \eqref{E:MetricAnsatz} corresponds to an energy momentum tensor
\begin{equation}
\label{E:Tmnstatic}
	\langle T_{\mu\nu} \rangle = \hbox{diagonal} \begin{pmatrix} \epsilon & \frac{1}{3} \epsilon & \frac{1}{3}\epsilon & \frac{1}{3} \epsilon \end{pmatrix}
\end{equation}
where 
\begin{equation}
\label{E:Gotepsilon}
	\epsilon = \frac{3}{2\kappa_5^2 b^4}\,,
\end{equation}
with $b$ given in \eqref{E:HawkingT}.

\subsection{The probe approximation}
\label{S:Probe}
At order $q^{-1}$, the equations of motion reduce to the Maxwell and Klein-Gordon equation for $\psi$ and $A_{m}$. We start with an ansatz where $A_0^{(1)}$ and $A_5^{(1)}$ are non zero, and the spatial components of the gauge field $A_i$ are turned on at the linear level.
Consider the gauge invariant variables:\footnote{%
Since we are working to leading order in $q$ in the matter fields, we can use $A_m$ and $\psi$ instead of $A_m^{(1)}$ and $\psi^{(1)}$. Alternately, we can omit the superscript $(1)$ from all expressions and reinsert them only when discussing boundary observables.
}
\begin{equation}
\label{E:xidef}
	G_{m}  = A_{m} - \partial_{m} \varphi\,.
\end{equation}
If we keep only the $m^2$ term in the scalar potential \eqref{E:scalarpotential},\footnote{In principle, one can carry out an appropriate rescaling of the couplings appearing in the scalar potential so that higher orders of $\psi$ will also contribute to the equations of motion at order $q^{-1}$. This has been done in, for example, \cite{Yarom:2009uq}.} 
we obtain the equations of motion
\begin{subequations}
\label{E:ProbeEOM}
\begin{align}
\label{E:xi0eom}
	\frac{d}{dr} \left(r^3 \frac{d G_0}{dr} \right) &=  \frac{b^4 r^5 \rho^2 G_0}{b^4 r^4-1} \\
\label{E:rhoeom}
	\frac{d^2 \rho}{d r^2} + \frac{5 b^4 r^4-1}{r(b^4 r^4-1)} \frac{d \rho}{dr} - \frac{b^4 m^2 r^2}{b^4 r^4-1}\rho & =  -\frac{b^8 r^4 \rho \; G_0^2}{(b^4r^4-1)^2} \,,
\end{align}
\end{subequations}
which are supplemented by the constraint
\begin{equation}
\label{E:Gotxir}
	G_5 = -\frac{b^4 r^2 G_0}{b^4 r^4-1}\,.
\end{equation}
In the more conventional Fefferman-Graham coordinate system it is customary to choose a gauge where the scalar field is real in which case one also finds that $A_5=0$. From \eqref{E:xidef} and \eqref{E:Gotxir} it is clear that in the Eddington-Finkelstein coordinate system one can not set both $\varphi=0$ and $A_5 = 0$ unless the scalar field vanishes entirely. Of course, once the solution in the Fefferman-Graham coordinate system is known, one can obtain the relation between $G_5$ and $G_0$ appearing in \eqref{E:Gotxir} by an appropriate coordinate transformation.

The boundary conditions we impose on our fields are that $G_{m}$ and $\rho$ are finite at the black hole horizon. Near the asymptotically AdS boundary we require that the source term for the scalar field $\psi_s$, defined in \eqref{E:psiexpansion}, vanish. From the constraint equation \eqref{E:Gotxir} we see that finiteness of $G_5$ at the horizon implies that $G_0$ must vanish there and finiteness of $G_5$ at the boundary follows from finiteness of $G_0$. Thus,
\begin{equation}
	\rho(1/b) = \hbox{\small finite} 
	\qquad
	G_{0}(1/b) = 0\,,
\end{equation}
and
\begin{equation}
	\rho \xrightarrow[r\to\infty]{} \mathcal{O}(r^{-\Delta})
	\qquad
	\lim_{r \to \infty} G_0 = \hbox{\small finite}\,.
\end{equation}

It is important to point out that we did not require, a priori, that $G_0$ vanish at the horizon. Rather, this restriction followed from the constraint equation \eqref{E:Gotxir}. When we consider a time-dependant (dynamical) superfluid in section \ref{S:Dynamical} then the constraint equation equivalent to \eqref{E:Gotxir} will imply that $G_0$ is non vanishing at the horizon. When working in the Fefferman-Graham coordinate system it is standard practice to require that $G_0$ (or $A_0$) vanish at the horizon on the grounds that $G_{0}dt$ have finite norm there. When working in ingoing Eddington-Finkelstein coordinates, a finite value of $G_0$ at $r=1/b$ will not generate a divergence on the future horizon and so, if there are no other constraints, we are free to keep its horizon value arbitrary. The norm of $G_0dt$ does diverge at the bifurcation point of the horizon if $G_0(1/b)\neq 0$. However, when constructing gradient corrections to the black hole geometry, the past horizon (which includes the bifurcation point) becomes singular so having $G_0$ diverge there is not a cause for worry. The singular nature of the past horizon is not surprising since generic solutions of viscous fluid dynamics are not expected to be regular in the infinite past.

Once a solution to \eqref{E:ProbeEOM} is obtained it is convenient to choose a gauge where $A_5=0$.
In this gauge, the phase of the scalar $\varphi$ is given by
\begin{equation}
\label{E:varphidef}
	\partial_r \varphi = -G_5 \qquad \partial_\mu \varphi = -\lim_{r \to \infty} G_{\mu}
\end{equation}
where the second term in \eqref{E:varphidef} comes from the requirement that $J_{\mu}$ is not sourced, by which we mean $\lim_{r \to \infty} A_{\mu} = 0$. Our definition of $\varphi$ allows for 
an arbitrary additive constant 
which has no physical significance.
With the gauge choice \eqref{E:varphidef}, the bulk to boundary identifications \eqref{E:goldstone} and \eqref{E:AToJ} take the form
\begin{equation}
\label{E:Gotdtphi}
	-\partial_t \langle \phi\rangle = \lim_{r \to \infty} G_0
\end{equation}
and
\begin{equation}
\label{E:GToJ}
	\kappa_5^2 \langle J_{\mu} \rangle  = \lim_{r \to \infty} r^2 G_{\mu} 
	+ \pdt\,.
\end{equation}
Using the definition of the chemical potential \eqref{E:Gotmu}, our solution for $\varphi$ \eqref{E:varphidef} and the horizon boundary condition $G_0(1/b)=0$, we find that \eqref{E:Gotdtphi} implies that
\begin{equation}
\label{E:GotJosephson}
	\partial_t \langle \phi \rangle = -\mu
\end{equation}
which is precisely the Josephson condition \eqref{E:Josephson} in the boundary theory in a configuration where the spatial components of the normal velocity vanish. Put differently, by requiring that $G_0$ and $G_5$ are finite at the future horizon and that $A_{0}$ is not sourced, the horizon asymptotics of the constraint equation \eqref{E:Gotxir} enforce the Josephson condition in the boundary theory.\footnote{%
  A numerical derivation of the Josephson condition 
  when $q$ is finite can be found in \cite{Sonner:2010yx}.
  }

By turning on $G_i$ we extend the solution to include a non vanishing superfluid flow $\partial_{i}\langle \phi \rangle \neq 0$ in the boundary theory. Working with a small value of $G_i$ corresponds in the current setup to setting $n^{\mu} \ll 1$ in the boundary theory. As  discussed in section \ref{S:Superfluids}, this limit is physically sensible since superfluidity breaks down at large superfluid velocities. The linearized equations for $G_i$ are given by
\begin{equation}
\label{E:xiiequation}
	\frac{d^2 G_i}{d r^2} + \frac{1+3 r^4 b^4}{r(b^4 r^4-1)} \frac{d G_i}{dr} = \frac{r^2 b^4 \rho^2}{r^4 b^4-1} G_i \,.
\end{equation}
Multiplying equation \eqref{E:xiiequation} by $G_0$ and using \eqref{E:xi0eom}, the equation for $G_i$ takes the form
\begin{equation}
\label{E:GiequationV2}
		\frac{d}{dr} \left( \frac{r^4 b^4-1}{4 r} \left( G_0 \frac{d G_i}{dr} - G_i \frac{d G_0}{dr} \right) \right)= -\frac{G_i}{r^2} \frac{d G_0}{dr}\,.
\end{equation}
We will find that the form \eqref{E:GiequationV2} is more useful than \eqref{E:xiiequation} when discussing the backreaction of the metric in section \ref{S:backreaction}. 
At the horizon we require that $G_i$ is finite. To understand what boundary conditions to impose on $G_i$ at the asymptotically AdS boundary we note that in a gauge where $A_5=0$, \eqref{E:varphidef} implies that
\begin{equation}
\label{E:boundaryxii}
	\partial_i \langle \phi \rangle = -\lim_{r \to \infty} G_i\,.
\end{equation}
Since \eqref{E:xiiequation} is linear, it is convenient to define $g(r)$ such that
\begin{equation}
\label{E:xdef}
	G_i = -g(r) \, \partial_i \langle \phi \rangle 
\end{equation}
where $\partial_i \langle \phi \rangle$ is a constant and $g$ satisfies the same equation as $G_i$ but with boundary conditions such that $g(\infty)=1$. In \cite{Herzog:2008he,Basu:2008st,Sonner:2010yx,Herzog:2010vz,Arean:2010wu} the equations of motion for $G_i$ were studied beyond the linearized approximation.

\subsection{The backreaction of the metric}
\label{S:backreaction}
At order $q^{-2}$ the metric gets corrected due to the matter fields. The most general isotropic order $q^{-2}$ gauge-fixed line element is
\begin{equation}
\label{E:2ndOrderMetric}
	g_{mn}^{(2)}dx^m dx^n = -r^2  f^{(2)}(r)dt^2 + 2 s^{(2)}(r) drdt\,.
\end{equation}
The equations of motion for $f^{(2)}$ and $s^{(2)}$ are
\begin{align}
\label{E:BREOM}
6 \frac{ds^{(2)}}{dr} & = \frac{b^8 r^5 \rho^2 G_0^2}{(b^4 r^4-1)^2} +  r \left(\frac{d\rho}{dr}\right)^2  \\
6 \frac{d}{dr}\left(r^4 f^{(2)}\right) -48 r^3 s & = \frac{b^4 r^5 \rho^2 G_0^2}{b^4 r^4-1}-m^2 r^3 \rho^2  - r^3\left(\frac{dG_0}{dr} \right)^2+ \frac{r(b^4 r^4 -1)}{b^4}\left(\frac{d\rho}{dr}\right)^2 \,.
\end{align}
We require that $s^{(2)}$ vanish at the asymptotically AdS boundary and that the solution for $f^{(2)}$ does not shift the mass of the Schwarzschild black hole $b$ in \eqref{E:MetricAnsatz},
\begin{equation}
	\lim_{r \to \infty} r^4 f^{(2)} = 0 + \pdt\,.
\end{equation}

Once we turn on $G_i$ at the linearized level, we must allow for fluctuations of the $g_{0i}^{(2)}=r^2 \gamma^{(2)}_i$ components of the metric  \cite{Herzog:2009md}.\footnote{%
   The $\gamma_i^{(2)}$ and $\gamma$ in this section are unrelated to the boundary metric $\gamma_{\mu\nu}$
   of section \ref{S:Static}.
 } 
 This coupling represents the physical fact that a charged current should carry momentum in a medium with nonzero charge density.
The  linearized Einstein equations for $\gamma_i^{(2)}$ are
\begin{equation}
\label{E:gammaiequation}
	\frac{d}{dr} \left(r^5 \frac{d \gamma^{(2)}_i}{dr} \right) = - r^3 \frac{d G_i}{dr} \frac{d G_0}{dr} - \frac{b^4 r^5 G_i \rho^2 G_0}{b^4 r^4-1}\,.
\end{equation}
Because the $\gamma_i^{(2)}$ are coupled linearly to the $G_i$'s, in analogy to \eqref{E:xdef} it is natural to define $\gamma$ through the relation
\begin{equation}
	\gamma_i^{(2)} \equiv - \gamma \, \partial_i \langle \phi \rangle \,.
\end{equation}
After a manipulation similar to the one that took us from \eqref{E:xiiequation} to \eqref{E:GiequationV2} we obtain
\begin{equation}
\label{E:gammaequation}
	\frac{d}{dr} \left(r^5 \frac{d \gamma}{dr} \right) = - \frac{d}{dr} \left(r^3 g \frac{d G_0}{dr} \right)\,.
\end{equation}
Using equation \eqref{E:GiequationV2} we find that the solution to \eqref{E:gammaequation} is
\begin{equation}
\label{E:gammasol}
	\gamma = \frac{r^4 b^4 - 1}{4 r} \left(G_0 \frac{d g}{dr} - g \frac{d G_0}{dr} \right) + \frac{\kappa^2 \rho_{\rm n}}{2 r^4} \left( b^4 r^4 - 1\right)\,.
\end{equation}
(The manipulations leading to \eqref{E:gammasol} essentially follow from the work of \cite{Gubser:2009qf}.)  The integration constants in \eqref{E:gammasol} were fixed as follows.
We require that $\gamma$ is finite at the horizon and that near the asymptotically AdS boundary it satisfies
\begin{equation}
\label{E:gammaboundary}
	\lim_{r \to \infty} r^4 \gamma = \frac{\kappa^2}{2} \rhos 	
							+ \left(\substack{\hbox{\tiny possible} \\ \hbox{\tiny divergent} \\ \hbox{\tiny terms}}\right)\,.
\end{equation}
The boundary condition \eqref{E:gammaboundary} follows from the linearized version of the inviscid expression for the energy momentum tensor \eqref{E:InviscidHydro},
\begin{align}
\label{E:T0ilinear}
\begin{split}
	\langle T_{0i} \rangle 
	&= (\epsilon+P-\mu \rhos)u_0 u_i + \frac{\rhos}{\mu} \partial_0\phi  \partial_i\phi \\
		  &\simeq -\rhos \partial_i \phi
\end{split}
\end{align}
and the relation \eqref{E:gmnToTmn} between the boundary theory energy momentum tensor and the bulk metric. The freedom we have in choosing the overall coefficient of the $r^{-4}$ term in a series expansion of $\gamma$ is due to possible shifts in the spatial components of the normal fluid velocity which we must set to zero.

It is interesting to note that the condition \eqref{E:gammaboundary} implies that the $g_{0i}$ components of the metric vanish at the horizon: 
\begin{equation}
\label{E:gammahzero}
	\gamma(1/b)=0\,. 
\end{equation}
If instead of \eqref{E:gammaboundary} we had used a different boundary condition for $\gamma$ then this would have implied, via \eqref{E:T0ilinear}, that we had turned on the spatial component of the normal velocity $u_i$. Once $u_i$ is non zero the spatial component of the entropy current defined in \eqref{E:SCurrent} will not vanish and we can expect from the analysis of \cite{Bhattacharyya:2008xc} that $g_{0i}(1/b) \neq 0$. Thus, requiring that $u_i=0$ implies $g_{0i}(1/b)=0$.

\subsection{A boosted static solution}
\label{S:Boosted}
The solution in sections \ref{S:AdSSS}-\ref{S:backreaction} is static in the sense that it is time-independent: the energy density and charge density are constant in spacetime and so is the velocity of the superfluid. It is straightforward to extend the static solution described so far to include a spacetime independent velocity field for the normal component. Let $\Lambda_{\mu\nu}$ denote a boost by a velocity parameter $-u^{\mu}$ where
\begin{equation}
\label{E:udef}
	u_{\mu} = \begin{pmatrix} -\frac{1}{\sqrt{1-\beta^2}} & \frac{\beta_1}{\sqrt{1-\beta^2}} & \frac{\beta_2}{\sqrt{1-\beta^2}} &  \frac{\beta_3}{\sqrt{1-\beta^2}} \end{pmatrix}
\end{equation}
and $\beta^2 = \sum \beta_i^2 < 1$.
Under the coordinate transformation $x^{\mu} \to x^{\prime\;\mu} = \Lambda^{\mu\nu}x_{\nu}$, the field $G_{\mu}$ transforms into
\begin{align}
\begin{split}
\label{E:Gprime}
	G_{\mu} ' &= \Lambda_{\mu}^{\nu} \begin{pmatrix} G_0 \\ -g \partial_i \phi \end{pmatrix}_{\nu} \\
		&= -G_0 u_{\mu} - g N_{\mu}
\end{split}
\end{align}
where we defined
\begin{equation}
	N^{\mu} \equiv \Lambda^{\mu i} \partial_i \langle \phi \rangle \,.
\end{equation}
The boosted metric takes the form
\begin{multline}
\label{E:BoostedBB}
	ds^{\prime\;2} = -r^2 \left( 1 - \frac{1}{(b^4 r^4)} + \frac{f^{(2)}}{q^2} \right) u_{\mu}u_{\nu} dx^{\mu}dx^{\nu} + r^2 P_{\mu\nu}dx^{\mu}dx^{\nu} -2 \left(1+\frac{s^{(2)}}{q^2}\right) u_{\mu}dx^{\mu}dr 
	\\
	+ 2 r^2 \gamma u_{(\mu} N_{\nu)} dx^{\mu} dx^{\nu}\,.
\end{multline}
In \eqref{E:BoostedBB} 
$P_{\mu\nu}$ is the projection matrix defined in \eqref{E:Projectiondef} and $s^{(2)}$ and $f^{(2)}$ are the order $q^{-2}$ corrections to the metric defined in \eqref{E:2ndOrderMetric}. The scalars $G_5$ and $\rho$ remain invariant under the coordinate transformation. We emphasize that \eqref{E:BoostedBB} and \eqref{E:Gprime} solve the equations of motion to order $q^{-2}$ since they have been obtained via a coordinate transformation of solutions.

In terms of boundary theory observables, we have
\begin{equation}
	\partial'_\mu \langle \phi \rangle= -\lim_{r \to \infty} G_\mu' 
\end{equation}
so that according to \eqref{E:Gprime}
\begin{align}
\begin{split}
\label{E:Nnu}
	 N_{\nu} &= -\lim_{r \to \infty} G_0 u_{\nu} + \partial'_{\nu} \langle \phi \rangle \\
	 	& =  -\mu u_{\nu}  + \partial^{\prime}_{\nu} \langle \phi \rangle \\
		& = \mu n_{\nu}\,
\end{split}
\end{align}
where in the second line we used \eqref{E:GotJosephson} and in the third line we used the fact that since the configuration we are considering is time independent $\mu_{\rm s} = \mu$. In the rest of this work we will consider only the boosted frame and will omit primes from boundary quantities which are all boosted. In the boosted frame the charged current is given by \eqref{E:InviscidJ} with 
\begin{align}
\begin{split}
\label{E:rhoandrhos}
	\kappa_5^2 \rhot &= -\lim_{r \to \infty}  r^2 G_0 +\pdt \\
	\kappa_5^2 \mu^{-1} \rhos & = -\lim_{r \to \infty} r^2 g +\pdt \,.
\end{split}
\end{align}
The energy momentum tensor in the boosted frame can be read off the metric \eqref{E:BoostedBB} using \eqref{E:gmnToTmn}. We find 
\begin{equation}
\label{E:TmnStatic}
	\langle T_{\mu\nu} \rangle = \epsilon u_{\mu} u_{\nu} + \frac{1}{3} \epsilon P_{\mu\nu} + 2 \rhos\mu u_{(\mu}n_{\nu)}
\end{equation}
where $\epsilon$ and $P_{\mu\nu}$ were defined in \eqref{E:Gotepsilon} and \eqref{E:Projectiondef}. This is the expected form for the inviscid energy momentum tensor \eqref{E:InviscidHydro} when working to linear order in $n^{\mu}$.

To summarize, our strategy for computing the bulk dual of a static holographic superfluid is to solve \eqref{E:ProbeEOM} and \eqref{E:BREOM} to obtain the matter fields $G_0$, $\rho$, $G_5$ and the subleading order corrections to the metric $s^{(2)}$ and $f^{(2)}$. With these expressions we can solve the linearized equations \eqref{E:xiiequation} and \eqref{E:gammaiequation} for the spatial component of the matter fields $G_i = -g \partial_i \langle \phi \rangle$ and the space-time component of the metric $G_{0i} = -r^2 \gamma \partial_i \langle \phi \rangle$.

\subsection{An explicit solution near $T_0$}
\label{S:Explicit}
In \cite{Herzog:2010vz} it was shown how to construct an explicit solution to \eqref{E:ProbeEOM}, \eqref{E:Gotxir} and \eqref{E:xiiequation} close to the phase transition. In what follows we reproduce the solution of \cite{Herzog:2010vz} and extend it to include the leading order backreaction of the metric. 

Working in the probe limit, we denote the solution to \eqref{E:ProbeEOM} in the normal phase by $\mu G_0^\oss{0}$ where according to \eqref{E:Gotmu} $\mu$ is the chemical potential of the boundary theory and
\begin{equation}
\label{E:xi0normal}
	G_0^\oss{0} = \left(1 - \frac{1}{b^2 r^2} \right)\,.
\end{equation}
Condensation of the scalar field implies that at a given temperature there exists a critical value of the chemical potential (which we denote by $\mu = \mu_0$) at which the equation of motion for $\rho$ in \eqref{E:rhoeom} admits a zero mode. By a zero mode we mean a solution to \eqref{E:rhoeom} but with $G_0$ replaced by $\mu_0 G_0^{\oss{0}}$. This solution, which we denote $\rho^\oss{1}$, is defined up to an overall multiplicative constant. Working perturbatively in the dimensionless quantity $(\mu - \mu_0) b$ it is possible to obtain the solution to \eqref{E:ProbeEOM} close to the critical chemical potential.\footnote{In \cite{Herzog:2010vz} a slightly different expansion was carried out where the small parameter was $\langle O_{\psi} \rangle$. The critical chemical potential can then be determined as a function of $\langle O_{\psi} \rangle$. See appendix \ref{A:Kubo} for details.} Such an expansion takes the form
\begin{align}
\begin{split}
\label{E:expansion}
	b G_0(r) &= \mu_0 b G^\oss{0}_0(br) + \sum_{n=1}^{\infty} G_0^\oss{2n}(br) (\mu b - \mu_0 b)^{n}
	\\
	\rho(r) &= \sum_{n=1}^{\infty} \rho^\oss{2n-1}(br) (\mu b-\mu_0 b)^{(2n-1)/2}\,.
\end{split}
\end{align}
The boundary condition $\lim_{r \to \infty} G_0 = \mu$ implies that $\lim_{r\to\infty}G_0^\oss{n} = 0$ for $n>2$ and $\lim_{r\to\infty}G_0^\oss{2} = 1$. We have conveniently defined the argument of $\rho^\oss{n}$ and $G_0^\oss{n}$ to be the dimensionless combination $br$. 

In practice the expansion \eqref{E:expansion} has been found useful only when the zero mode of $\rho$ can be obtained in closed form. 
Such a closed form solution exists for the special case
 $m^2 = -4$,\footnote{%
 The case $m^2=-4$ corresponds to $\Delta=2$ where the series expansion \eqref{E:psiexpansion} reads 
\begin{equation*}
		\psi = \psi_s r^{-2}\ln r \left(1 + \ldots\right)  + \psi_r {r^{-2}} \left(1 + \ldots\right)\,.
\end{equation*}
} 
where the phase transition occurs at
\begin{equation}
\mu_0 b = 2 \ .
\end{equation}
Plugging the expansion \eqref{E:expansion} into the equations of motion \eqref{E:ProbeEOM} and expanding in powers of $\mu b - \mu_0 b$ 
one finds the following equations of motion for $G_0^{\oss{n}}(x)$ and $\rho^{\oss{n}}(x)$:
\begin{eqnarray}
\label{E:GotGn}
	\left(x^3 G_0^{\oss{n}\;\prime}(x) \right)' &=& S_0^\oss{n}(x) \ ,
\\
\left(\frac{x(x^2-1)}{x^2+1} \left((x^2+1) \rho^{\oss{n}}(x) \right)' \right)' &=& S_{\rho}^{\oss{n}} \ ,
\end{eqnarray}
where $S_0^\oss{n}$ and $S_{\rho}^{\oss{n}}$ are functions of the lower order solutions 
$\rho^\oss{m}$ and $G_0^\oss{m}$ with $m<n$.
The first few solutions for $G_0^\oss{n}$ and $\rho^\oss{n}$ are given in appendix 
\ref{A:exactsols}.

Once the solution to \eqref{E:ProbeEOM} has been obtained perturbatively, it is a simple exercise to compute the solutions to the linearized vector equations for $G_i$, \eqref{E:xiiequation}. In the notation of \eqref{E:xdef} we expand the vector modes near $\mu_0 b =2$ such that
\begin{equation}
\label{E:gexpansion}
	g(r) = \sum_{n=0}^{\infty} g^\oss{2n}(r b) (\mu b-2)^{n} \,.
\end{equation}
Imposing the boundary condition $g(\infty)=1$ amounts to setting $\lim_{x \to \infty} g^{\oss{n}} = \delta^{\overline{n}0}$. Inserting the expansion \eqref{E:gexpansion} 
into \eqref{E:xiiequation}, we find that the equation of motion for $g^{\oss{n}}$ takes the form
\begin{equation}
\label{E:EOMg}
	\left(\frac{x^4-1}{x} g^{\oss{n}\prime}(x)\right)' = S_{g}^{\oss{n}}\,.
\end{equation}
where $S_g^{\oss{n}}$ depends on $g^{\oss{m}}$, $G_0^{\oss{m}}$ and $\rho^{\oss{m}}$
for $m < n$.
The first few values of $g^{\oss{n}}$ can be found in appendix \ref{A:exactsols}.

Using  the explicit solutions for $g^{\oss{n}}$, $G_0^{\oss{n}}$ and $\rho^{\oss{n}}$
and the bulk to boundary mapping described in section \ref{S:Static} and summarized in appendix \ref{A:BtoB}, we can compute $| \langle O_{\psi} \rangle |$ and the boundary theory current $\langle J_{\mu} \rangle$ 
 for a configuration moving with a superfluid velocity proportional to  $\partial^{\mu} \langle \phi \rangle$. More explicitly, using \eqref{E:GToJ}, \eqref{E:GotOpsi} and \eqref{E:InviscidHydro} with $u^{\mu}=(1,\vec{0})$, we find that
\begin{align}
\begin{split}
	\kappa_5^2 b^3 \rhot &= 2 + 7 (\mu b - 2) - \frac{3}{4} (96 \ln 2-71) (\mu b - 2)^2 + \mathcal{O}\left((\mu b-2)^3\right)\\
	\kappa_5^2 b^2 \mu^{-1} \rho_s &= 6 (\mu b - 2) - \left(96\ln 2-\frac{241}{4}\right) (\mu b - 2)^2 + \mathcal{O}\left((\mu b-2)^3\right)
\end{split}
\label{E:rhoandrhosseries}
\end{align}
and
\begin{equation}
\label{E:Oval}
	\frac{\kappa_5^2 b^2}{C_{\Delta}} | \langle O_{\psi} \rangle | = 4 \sqrt{3} (\mu b - 2)^{1/2} + \frac{253-336\ln 2}{4 \sqrt{3}} (\mu b - 2)^{3/2} + \mathcal{O}\left( (\mu b - 2)^{5/2} \right)\,.
\end{equation}
Similar expressions can be found in \cite{Herzog:2010vz}.

The equations of motion for the backreacted metric are similar to those of the matter fields. Expanding the corrections to the metric $s^{(2)}$, $f^{(2)}$ and $\gamma$,
\begin{align}
\begin{split}
	s^{(2)} &= \sum_{n=0}^{\infty} s^{(2,\overline{2n})}(r b) (\mu b-2)^{n} \\
	f^{(2)} &= \sum_{n=0}^{\infty} f^{(2,\overline{2n})}(r b) (\mu b-2)^{n} \\
	\gamma &= b\sum_{n=0}^{\infty} \gamma^\oss{2n}(r b) (\mu b-2)^{n}
\end{split}
\end{align}
and inserting them into the equations of motion \eqref{E:BREOM} and into \eqref{E:gammasol}, we obtain a set of equations for $s^{(2,\overline{n})}$, $f^{(2,\overline{n})}$ and $\gamma^\oss{n}$. The solutions for the first few $\oss{n}$ are given in appendix \ref{A:exactsols}.

\section{Dynamical holographic superfluids}
\label{S:Dynamical}
To generate a spacetime dependent holographic superfluid we can follow the same strategy that led us from the inviscid superfluid in section \ref{S:Inviscid} to the viscous superfluid in section \ref{S:Viscous}: we allow the hydrodynamic variables to depend on the spacetime coordinates and look for the appropriate corrections to the bulk fields. This is essentially the strategy used by \cite{Bhattacharyya:2008jc} to construct the metric dual to a viscous fluid. In section \ref{S:GEPO0} we work out the dual of viscous flow to leading order in the large $q$ limit. Since at leading order only the metric enters into the equations of motion, this is essentially a rederivation of the results of \cite{Bhattacharyya:2008jc}. The reader familiar with the work of \cite{Bhattacharyya:2008jc} may go directly to section \ref{S:GEPO1} where we compute the effects of viscosity on the bulk matter fields. In section \ref{S:GEPO2} we argue that the shear viscosity associated with the superfluid motion vanishes. Finally, in section \ref{S:ExplicitO1} we obtain an explicit expression for the third bulk viscosity, diffusion coefficient and viscous corrections to the order parameter close to the phase transition using the solution in \ref{S:Explicit}.

\subsection{Viscous flow from the Schwarzschild black hole}
\label{S:GEPO0}
Consider the solution \eqref{E:BoostedBB} to the Einstein equations but with $s^{(2)}=f^{(2)}=\gamma=0$, i.e., the boosted AdS-Schwarzschild black hole. The boundary theory stress tensor dual to the configuration in \eqref{E:BoostedBB} is given by \eqref{E:TmnStatic} with $\rhos=0$ which describes an inviscid fluid moving at uniform velocity $u^{\mu}$. Viscous corrections to the fluid motion vanish since gradients of the velocity field and energy density are zero.

To obtain a configuration where the velocity field and energy density are not uniform we promote $\beta_i$ in \eqref{E:udef} and the inverse temperature $b$ in \eqref{E:TmnAnsatz} to be spacetime dependent. In doing so \eqref{E:BoostedBB} is no longer a solution to the Einstein equations and the metric has to be corrected. We denote the correction of the metric by $\delta g^{(0)}_{mn}$ and, following \cite{Bhattacharyya:2008ji}, parametrize it by:
\begin{multline}
\label{E:deltametric}
	\delta g_{mn}^{(0)}  dx^{m}dx^{n} = - 2  u_{\mu} dx^{\mu} r \left( u^{\alpha}\partial_{\alpha}u_{\nu} + \frac{1}{3} \partial_{\alpha}u^{\alpha} u_{\nu} \right) dx^{\nu} -2 \delta s^{(0)}  u_{\mu} dx^{\mu}  dr   
	+ r^2 \delta h^{(0)} P_{\mu\nu} dx^{\mu} dx^{\nu} 
	\\
	- r^2 \delta f^{(0)} u_{\mu} dx^{\mu} u_{\nu} dx^{\nu} + r^2 2 \delta V^{(0)}_{(\mu} u_{\nu)} dx^{\mu}dx^{\nu} + r^2 \delta \pi^{(0)}_{\mu\nu} dx^{\mu}dx^{\nu}\,.
\end{multline}
Here $\delta \pi^{(0)}$ is symmetric and traceless and both $\delta V^{(0)}$ and $\delta \pi^{(0)}$ are orthogonal to the velocity field $u^{\mu}$. Circular parentheses were defined in \eqref{E:SymmProduct} and denote a symmetric combination. We will fix most of our gauge freedom by setting $\delta h^{(0)}=0$ as in \cite{Haack:2008cp}. At this point it would perhaps be useful to recapitulate our notation. In section \ref{S:Static} we used an unbarred superscript to denote coefficients in a series expansion in the inverse charge $1/q$. In section \ref{S:Explicit} we used a barred superscript to denote coefficients in a series expansion in the chemical potential. Now we use a $\delta$ to denote gradient corrections to the metric. Our notation differs from the ones used in \cite{Bhattacharyya:2008jc,Haack:2008cp,Bhattacharyya:2008mz} where a superscript denoted coefficients in a gradient expansion and from the one in \cite{Haack:2008cp} where a barred superscript denoted coefficients in a near boundary expansion.

We will not attempt to solve the Einstein equations entirely. Rather, we neglect all terms which involve two derivatives of the velocity field and inverse temperature and look for a solution linear in these gradients. We will denote the terms that have been neglected by $\mathcal{O}(\partial^2)$. This means that the metric components in \eqref{E:deltametric} depend only on one derivative of $b$ or $u^{\mu}$. The equations of motion for $\delta s^{(0)}$,  $\delta f^{(0)}$, $\delta V^{(0)}$ and $\delta \pi^{(0)}$ naturally decompose themselves into scalar, tensor and vector equations under the $SO(3) \subset SO(3,1)$ under which $u_{\mu}$ is invariant. They are given by
\begin{align}
\label{E:GEO0equations}
\begin{split}
	\frac{d}{dr} \delta s^{(0)} &= 0 \\
	\frac{d}{dr} \left( r^4 \delta f^{(0)} \right) - 8 r^3 \delta s^{(0)} & = -4 r^2 \partial_{\alpha}u^{\alpha} \\
	\frac{d}{dr}\left( r^5 \frac{d \delta V^{(0)}_{\mu}}{dr} \right) &= 0 \\
	\frac{d}{dr} \left(r (r^4 b^4 -1) \frac{d \delta \pi^{(0)}_{\mu\nu}}{dr} \right) &= 3 r^2 b^4 \sigma^n_{\mu\nu} 
\end{split}
\end{align}
with $\sigma^n_{\mu\nu}$ defined in \eqref{E:sigmas}.
In addition to the equations of motion \eqref{E:GEO0equations} there are four constraint equations
\begin{equation}
\label{E:constraints}
	 u^{\mu} \partial_{\mu} b = \frac{1}{3} b \partial_{\mu} u^{\mu} 
	 \qquad
	 P^{\mu\nu}\partial_{\nu} b = b u^{\nu} \partial_{\nu} u^{\mu}\,.
\end{equation}
(Equations \eqref{E:constraints} amount to four independent equations since the matrix $P_{\mu\nu}$ is a projection.) 
The constraint equations are equivalent to energy momentum conservation $\partial_{\mu}T^{\mu\nu} = 0$ with an energy momentum tensor as in \eqref{E:TmnStatic} but with $\rhos$ set to zero.
The boundary conditions we impose on the metric components are that they are not singular at the horizon which is located at $r=1/b$ and that they do not deform the boundary theory metric. The requirement that $u^\mu u^\nu T_{\mu\nu} = \epsilon$ sets the $\mathcal{O}(r^{-4})$ term of the near boundary series expansion of $\delta f^{(0)}$ to zero.  Working in the Landau frame \eqref{E:DefLandau} sets the $\mathcal{O}(r^{-4})$ term of $\delta V^{(0)}$ to zero.  
We refer the reader to \cite{Bhattacharyya:2008jc,Haack:2008cp} for more details. 

The solution to \eqref{E:GEO0equations} is given by
\begin{align}
\begin{split}
	\delta s^{(0)} & = 0 \ , \; \; \;
	\delta f^{(0)}  = \frac{4}{3 r} \partial^{\alpha}u_{\alpha} \ , \; \; \;
	\delta V^{(0)}_{\mu}  = 0 \ ,  \\
	b^{-1} \delta \pi^{(0)}_{\mu\nu} & = \left(\frac{\pi}{4} - \frac{1}{2} \arctan\left(r b \right) - \ln (r b) + \frac{1}{2}\ln(1+r b) + \frac{1}{4} \ln\left(1+r^2 b^2\right) \right) \sigma_{\mu\nu}^n\,.
\end{split}
\end{align}
Following \eqref{E:gmnToTmn} the boundary stress tensor is given by
\begin{equation}
	2\kappa_5^2 \langle T_{\mu\nu} \rangle =\frac{1}{b^4} \left(4 u_{\mu}u_{\nu} + \eta_{\mu\nu}  \right) - \frac{1}{b^3} \sigma_{\mu\nu}^n.
\end{equation}
Using the Bekenstein-Hawking entropy formula we obtain
\begin{equation}
	s = \frac{2 \pi b^3}{\kappa_5^2}
\end{equation}
which gives us the celebrated relation \eqref{E:Universal}.

For the reader unfamiliar with the formalism introduced in \cite{Bhattacharyya:2008jc}, we point out that a key feature of the equations of motion which allows one to simplify them considerably is that they are ultra-local: one can solve equations \eqref{E:GEO0equations} in a neighborhood of a point $x_0^{\mu}$ and patch these solutions together. Technically, this feature of the equations of motion allows one to work in the neighborhood of a point $x_0^{\mu}$ where the velocity field and inverse temperature are chosen to satisfy
\begin{align}
\begin{split}
\label{E:atpoint}
	u^{\nu} &= \begin{pmatrix} 1 & 0 & 0 & 0 \end{pmatrix} + x^{\mu} \partial_{\mu} u^{\nu} + \mathcal{O}(\partial^2)  \\
	b & = b_0 + x^{\mu}\partial_{\mu} b + \mathcal{O}(\partial^2)\,.
\end{split}
\end{align}
When working with a computational software program such as \verb+Mathematica+ the decomposition \eqref{E:atpoint} significantly simplifies computations. We refer the reader to \cite{Bhattacharyya:2008jc} for details.

\subsection{Viscous superfluid hydrodynamics in the probe limit}
\label{S:GEPO1}
To compute the viscous corrections to the current and to the Josephson condition we follow the same strategy as the one outlined in section \ref{S:GEPO0}. In addition to promoting $b$ and $u^{\mu}$ to spacetime dependent quantities, we also allow the chemical potential $\mu$ and the field  $N^{\mu}$ (of which we keep only linear terms) to be spacetime dependent. Eventually we will be interested in configurations where $n^{\mu}$ is small so we will set $N^{\mu}$ to zero but not its derivatives.\footnote{%
To be precise, going back to \eqref{E:Nnu}, $N_{\mu}$ receives viscous corrections because $\mu n^{\mu} \neq -\mu u^{\mu} + \partial'_{\mu} \langle \phi\rangle $. Since $n_{\mu}$ is small and we are working to linear order in gradients these corrections do not play a role in our computations.
} 
We then look for corrections to $G_m$, $\rho$ and $\varphi$ which we denote by $\delta G_m$, $\delta\rho$ and $\delta\varphi$. It is convenient to decompose $\delta G_\mu$ into terms parallel and orthogonal to the normal velocity $u_{\mu}$:
\begin{equation}
	\delta G_{\mu} = -\delta G u_{\mu} + \delta g_{\mu} 
\end{equation}
with $u^{\mu} \delta g_{\mu} = 0$. We then require that the Maxwell and Klein-Gordon equations of motion be satisfied to first order in gradients of the hydrodynamic variables.

As in the case of the Schwarzschild black hole, we can use the ultralocal nature of the equations for $\delta \rho$, $\delta G$ and $\delta g_{\mu}$ to solve the equations of motion around a point $x_0^{\mu}=0$ where the fields $X = (G_0, G_5, \rho)$ can be expanded in the form
\begin{align}
\label{E:LocalExpansion}
\begin{split}
X(r; b(x^{\mu}), \mu(x^{\mu})) &= X(r;b(0),\mu(0)) + \frac{\partial X}{\partial b}\Bigg|_{x=0} x^{\nu}\partial_{\nu} b + \frac{\partial X}{\partial \mu}\Bigg|_{x=0} x^{\nu}\partial_{\nu} \mu +\mathcal{O}(\partial^2)  \ .
\end{split}
\end{align}
Because we work in a limit where $n^\mu$ is small, the expansion for $G^i$ is simpler:
\begin{align}
\begin{split}
	G_{i}(r; b(x^{\mu}), \mu(x^{\mu}), \partial_i \phi(x^{\mu})) &= -g(r;b(x^{\mu}), \mu(x^{\mu})) \partial_\alpha\phi(x^{\mu}) =  -g(r;b(0),\mu(0)) x^{\nu} \partial_{\nu} \partial_\alpha \phi + \mathcal{O}(\partial^2)\,.
\end{split}
\end{align}
The transformation properties of $\delta G$, $\delta g_{\mu}$, $\delta G_5$ and $\delta \rho$ under the $SO(3) \subset SO(3,1)$ under which $u^{\mu}$ is invariant imply that the equations of motion for the scalars $\delta G$, $\delta G_5$ and $\rho$ will decouple from the equations of motion for the vector $\delta g_{\mu}$.

\subsubsection{Scalar sector}
In the scalar sector the equation of motion for $\delta G$ and $\delta\rho$ are
\begin{align}
\label{E:deltaGEOM}
\begin{split}
	\frac{b^4 r^4-1}{b^4 r^5} \frac{d}{dr} \left(r^3 \frac{d}{dr} \delta G\right) = &
	\delta G \rho^2 + 2 \rho G \delta \rho + \frac{5-r^4 b^4}{r(r^4 b^4-1)} \mathcal{D}G_0 - 2 \frac{d}{dr} \mathcal{D}G_0 \\
	&+ \frac{\partial_{\alpha}u^{\alpha}}{3(r^4 b^4-1)}
		\left( -\frac{4(3 r^4 b^4+5)}{r(r^4b^4-1)}G_0 + 2 r^3 b^4 \rho^2 G_0 - (3 r^4 b^4-7) \frac{d}{dr}G_0 \right)
\end{split}
\end{align}
and
\begin{align}
\begin{split}
\label{E:deltarhoEOM}
	\frac{r^8b^8+1 - 2 r^4 b^4}{r^4b^8} &\left(\frac{d^2}{d r^2}\delta\rho + \frac{5 b^4 r^4-1}{r(b^4 r^4-1)} \frac{d }{dr}\delta\rho - \frac{b^4 m^2 r^2}{b^4 r^4-1}\delta\rho \right) =
		\\
		&-2 \rho G \delta G - G^2 \delta \rho 
		-\frac{r^4 b^4-1}{r^3 b^4} \left(3 \mathcal{D}\rho +2 r \frac{d}{dr} \mathcal{D}\rho \right)
		\\
		&+\frac{1}{3} \partial_{\alpha}u^{\alpha}  \left( 2 m^2 r \rho - \frac{4 r^3 b^4 \rho G_0^2}{b^4 r^4-1} - \frac{5 r^4 b^4 + 3}{r^2 b^4} \frac{d}{dr}\rho\right)\,.
\end{split}
\end{align}
In both \eqref{E:deltaGEOM} and \eqref{E:deltarhoEOM} we have used
\begin{equation}
\label{E:defD}
	\mathcal{D} X = \frac{\partial X}{\partial \mu} u^{\alpha}\partial_{\alpha}\mu + \frac{b}{3}  \frac{\partial X}{\partial b} \partial_{\alpha}u^{\alpha}
\end{equation}
with $X$ a bulk field.
The constraint equation is given by
\begin{align}
\begin{split}
\label{E:deltaG5}
	\delta G_5 =& -\frac{r^2 b^4 \delta G}{r^4 b^4 -1} + \frac{\partial_{\mu}N^{\mu}}{r^2 \rho^2} \frac{dg}{dr}
		+ \frac{r^2 b^4}{(r^4 b^4-1)\rho^2} \frac{d}{dr}\mathcal{D}G_0  \\
		&+\frac{\partial_{\nu}u^{\nu} b^4r^2}{r^4b^4-1}   \left(\frac{1}{\rho^2}\frac{dG_0}{dr} - \frac{2 r^3 b^4 G_0}{3(r^4b^4-1)} \right)\,.
\end{split}
\end{align}
In presenting \eqref{E:deltaGEOM}-\eqref{E:deltaG5} we have used \eqref{E:constraints}.

The boundary conditions on $\delta \rho$ and $\delta G$ are that they are finite at the horizon and that the source term $\psi_s$ in \eqref{E:psiexpansion} vanishes. We impose
\begin{equation}
\label{E:deltaGBoundary}
	\lim_{r \to \infty} r^2 \delta G = 0 + \pdt
\end{equation}
which amounts to the requirement that $u^{\mu}J_{\mu} = -\rhot$. To the order we are working in, we can also consistently set
\begin{equation}
\label{E:GoldstoneViscous}
	 \lim_{r \to \infty}  \left(G'_{\mu} + \delta G_{\mu}\right) = \lim_{r \to \infty} \left(-G_{0}u_{\mu} - g N_{\mu} + \delta G_{\mu}\right)  = -\partial_{\mu} \phi\,.
\end{equation}
The viscous corrections to $\partial_{\mu} \phi$ are encoded in the boundary value of $\delta G_{\mu}$. By contracting \eqref{E:GoldstoneViscous} with $u^{\mu}$ we obtain an expression for the viscous corrections to the Josephson condition,
\begin{equation}
\label{E:JosephsonCorrection}
	 -u^{\mu} \partial_{\mu} \phi = \lim_{r \to \infty} \left( G_0 + \delta G \right) \,.
\end{equation}
Here, as in \eqref{E:Gprime} $G'_{\mu}$ denotes the value of $G_{\mu}$ in the boosted frame. In principle, since in this section the velocity of the normal component is space-time dependent and generically does not vanish, we could have used primed variables such as $\delta G_{\mu}'$ instead of the current ones. We have not done so in order to avoid cluttering the notation. 

Once we take $n^{\alpha}$ to be small, the near boundary asymptotic expansion of $\delta G_5$ takes the form
\begin{equation}
\label{E:dG5boundary}
	\delta G_5 = \frac{2 C_{\Delta} r^{2\Delta-5}}{|\langle O_{\psi} \rangle|^2 \kappa_5^4} \partial_{\alpha} \left( n^{\alpha} \rhos + u^{\alpha} \rhot \right) + \mathcal{O}(r^{2\Delta-6})\,.
\end{equation}
Finiteness of the action implies that $\delta G_5$ should fall off faster than $r^{-2}$ near the boundary. One way to see this is to consider a gauge where $\delta\varphi=0$. In this gauge, the boundary value of the variation of the action with respect to $A_{\mu}$ would receive a divergent contribution from $\delta G_5$ unless it falls off fast enough at large $r$. Thus, \eqref{E:dG5boundary} together with \eqref{E:InviscidHydro} imply current conservation in the boundary theory.

The constraint equation \eqref{E:deltaG5} also supplies us with the correct horizon value of $\delta G$. Near the horizon, the constraint equation \eqref{E:deltaG5} together with regularity of $\delta\varphi$ implies that
\begin{equation}
\label{E:HConstraint}
	\delta G(1/b) = -\frac{1}{12} \frac{d G_0}{dr}\Bigg|_{r b= 1} \partial_{\nu}u^{\nu} + \frac{1}{\rho^2} \mathcal{D} \frac{dG_0}{dr} \Bigg|_{rb=1}\,.
\end{equation}
To obtain \eqref{E:HConstraint} we expanded $G_0$ and $\rho$ near the horizon located at $r b =1$. As opposed to the static configuration where we had $G_0(1/b)=0$, here $\delta G$ does not vanish at the event horizon. Since $\delta G (1/b) \propto \rho(1/b)^{-2}$ one might erroneously conclude that $\delta G(1/b)$ diverges in the $\mu \to \mu_0$ limit where $\rho$ vanishes. To see that this is not the case consider the solution to the equations of motion near the phase transition. As discussed in section \ref{S:Explicit} the solution takes the form
\begin{equation}
\label{E:G0nearTc}
	G_0  = \mu \left(1-\frac{1}{r^2 b^2}\right) + \mathcal{O}\left( \mu b-\mu_0 b \right)
\end{equation}
and
\begin{equation}
\label{E:rhonearTc}
	\rho = \mathcal{O}\left( (\mu b - \mu_0 b)^{1/2}\right)\,
\end{equation}
for any value of the mass of the scalar field.
Inserting \eqref{E:G0nearTc} and \eqref{E:rhonearTc} into \eqref{E:HConstraint} we find that
\begin{equation}
	\delta G(1/b) \propto \frac{\mu_0 \partial_{\mu} u^{\mu} + 3 u^{\mu} \partial_{\mu} \mu}{\mu b - \mu_0 b}\,.
\end{equation}
To leading order, current conservation close to the phase transition amounts to
\begin{align}
\begin{split}
	0 & = \partial_{\mu} \left( \rho u^{\mu} \right)\\
		&= \frac{\partial \rho}{\partial b} u^{\nu} \partial_{\nu}b + \frac{\partial \rho}{\partial \mu} u^{\nu} \partial_{\nu}\mu + \rho \partial^{\nu}u_{\nu}  \\
		&\simeq -\frac{2\rho}{b} u^{\nu} \partial_{\nu}b + \frac{\rho}{\mu} u^{\nu} \partial_{\nu}\mu + \rho \partial^{\nu} u_{\nu} \\
		& \simeq \left(\frac{1}{3} \partial^{\nu}u_{\nu} + \frac{1}{\mu_0} u^{\nu} \partial_{\nu}\mu \right) \rho\,.
\label{E:Leadingcurrent}
\end{split}
\end{align}
In the third line we used the fact that near the phase transition, $\rho \propto b^{-2}$ (which follows from $\rho \propto \mu$ and dimensional analysis). In the last line we have used \eqref{E:constraints}.
Thus, close to the phase transition the leading divergent contribution to $\delta G$ vanishes and it follows that
\begin{equation}
\label{E:dG0power}
	\delta G(1/b) = \mathcal{O}\left(  (\mu b - \mu_0 b)^0 \right)\,.
\end{equation}

Without an explicit solution to the equations of motion \eqref{E:ProbeEOM}, \eqref{E:deltarhoEOM} and \eqref{E:deltaGEOM} it is difficult to compute the viscous corrections to the Josephson condition and to the expectation value of $\mathcal{O}_{\psi}$. However, it is possible to argue that $\zeta_3$ will diverge close to the phase transition like the inverse power of the order parameter squared, or $(\mu b - \mu_0 b)^{-1}$. 
Equation \eqref{E:deltaGEOM} together with \eqref{E:dG0power} suggest that
$\delta G$ should scale as $(\mu b - \mu_0 b)^0$.
Thus, we should find that close to the phase transition,
\begin{equation}
\label{E:JosephsonScaling}
	u^{\mu} \partial_{\mu}\phi = -\mu  +  \mathcal{O} \left( (\mu b- \mu_0 b)^0 \right)\,.
\end{equation}
Comparing \eqref{E:JosephsonScaling} with \eqref{E:Josephson} and taking into account that $\rhos \sim \mathcal{O}\left( \mu b - \mu_0 b \right)$ we conclude that $\zeta_3 \sim \mathcal{O} \left( ( \mu b-\mu_0 b)^{-1} \right)$.

\subsubsection{Vector sector}
\label{S:Vsector}
The equations of motion for the vector modes read
\begin{multline}
\label{E:VectorModesO1}
	\frac{1}{r b^4} \frac{d}{dr} \left(\frac{r^4 b^4 -1}{r} \frac{d \delta g_\mu}{dr} \right) =
	\delta g_{\mu} \rho^2 +
	\\
	+ \frac{u^{\alpha}\partial_{\alpha}u_{\mu}}{r(r^4 b^4-1)^2} \left( \left( r^8 b^8 - 14 r^4 b^4 -3\right) G_0 + 2 r\left (-1+r^8 b^8\right) \frac{d G_0}{dr} \right) 
	+u^{\alpha} \partial_{\alpha} N_{\mu} \left( \frac{g}{r} + 2 \frac{d g}{dr} \right)\,.
\end{multline}
The boundary conditions we impose on $\delta g_{\mu}$ are that they are finite at the horizon and that $N^{\mu}$ doesn't get corrected at the boundary, i.e. $\lim_{r \to \infty} g_{\mu} = 0$.\footnote{Had we been interested in second order transport coefficients we would have needed to set $\lim_{r \to \infty} g_{\mu} \neq 0$. See \cite{Bhattacharyya:2008jc} for details.}  Note that using \eqref{E:phiSymmetries} and \eqref{E:HawkingT} we find that
\begin{equation}
	u^{\alpha} \partial_{\alpha}  N_{\mu} = -\frac{1}{\pi b} P_{\mu}^{\phantom{\mu}\nu} \partial_{\nu} \frac{\mu}{T} \,.
\end{equation}

As was the case for the scalar modes, it is difficult to solve \eqref{E:VectorModesO1} explicitly. In what follows we will study the behavior of $\delta g_{\mu}$ close to the phase transition and argue that the diffusion coefficient $\kappa$ defined in \eqref{E:ViscousCorrections} is continuous across $\mu_0$. Close to the phase transition we can use \eqref{E:G0nearTc} and \eqref{E:rhonearTc}  together with
\begin{equation}
\label{E:gmunearTc}
	g = 1+ \mathcal{O}\left(  \mu b - \mu_0 b \right)
\end{equation}
to solve \eqref{E:VectorModesO1} for $\delta g_{\mu}$. We find that the leading order solution to $\delta g_{\mu}$ is
\begin{multline}
	\delta g_{\mu} = \frac{1}{4\pi} \left( \pi - 2 \arctan(r b) + 2 \ln(1+r b) - \ln(1+r^2 b^2)\right) P_{\mu}^{\phantom{\mu}\nu} \partial_{\nu} \frac{\mu}{T}
	\\
	-\frac{\mu_0 b^2 r}{r^2 b^2 + 1} u^{\alpha} \partial_{\alpha} u_{\mu}
	+\mathcal{O}\left((\mu b - \mu_0 b)\right)\,.
\end{multline}
Using 
\begin{equation}
\label{E:JVEV}
	\kappa_5^2 \langle J_{\mu} \rangle = \lim_{r \to \infty} r^2 \left(G_{\mu} +  \delta G_{\mu}\right) + \left(\substack{  \hbox{\tiny divergent} \\ \hbox{\tiny terms}}\right)\,,
\end{equation}
we find that $\Upsilon_{\mu}$, the viscous correction to $J_{\mu}$, receives a contribution of the form
\begin{equation}
	\Upsilon_{\mu} = -\frac{1}{2 \pi b^2} P_{\mu}^{\phantom{\mu}\nu} \partial_{\nu} \frac{\mu}{T} + \mathcal{O}\left(\mu b - \mu_0 b\right)\,,
\end{equation}
from which we conclude that
\begin{equation}
	\kappa = \frac{\pi T^2}{2 \kappa_5^2} + \mathcal{O}\left(\mu b - \mu_0 b \right)\,.
\end{equation}
The leading term for $\kappa$ agrees with the value obtained in the normal phase when the charge of the black hole is taken to be small \cite{Son:2006em,Erdmenger:2008rm}. Thus, the diffusion coefficient $\kappa$ is continuous across the phase transition. 

\subsection{Viscous superfluids at order $q^{-2}$}
\label{S:GEPO2}
Our analysis of superfluids in section \ref{S:Viscous} revealed that within our approximation, the stress tensor has two possible shear viscosities: the familiar shear viscosity, $\eta$, associated with a $\partial_{\langle \mu} u_{\nu \rangle}$ term and a shear viscosity, $\eta_s$, associated with the gradients of the relative superfluid velocity $\partial_{\langle \mu} n_{\nu \rangle}$. Standard arguments due to Landau \cite{Landau} which we presented in \ref{S:Viscous} show that $\eta_s = 0$. These arguments rely on an assumption of the form of the entropy current and have been known to fail in certain subtle cases \cite{Erdmenger:2008rm,Banerjee:2008th,Son:2009tf}. The goal of this section is to compute $\eta_s$ and $\eta$ holographically. We will show that $\eta_s=0$ as expected, and that the shear viscosity of the normal component satisfies the universal relation \eqref{E:Universal} as was already discussed in \cite{Natsuume:2010ky}.

As in section \ref{S:GEPO0} we begin by decomposing the order $q^{-2}$ corrections to the metric into scalars $\delta s^{(2)}$, $\delta f^{(2)}$, a vector $\delta V^{(2)}_{\mu}$ satisfying $u^{\mu} \delta V^{(2)}_{\mu} = 0$ and a traceless symmetric tensor $\delta \pi^{(2)}_{\mu\nu}$ which is orthogonal to the normal velocity $u^{\mu} \delta \pi^{(2)}_{\mu\nu} = 0$:
\begin{multline}
	\delta g_{mn}^{(2)} dx^{m}dx^{n} = -2 \delta s^{(2)}  u_{\mu} dx^{\mu}  dr   \\
	 - r^2 \delta f^{(2)} u_{\mu} dx^{\mu} u_{\nu} dx^{\nu} + 2 r^2 \delta V^{(2)}_{(\mu}u_{\nu)} dx^{\mu}dx^{\nu} + r^2 \delta \pi^{(2)}_{\mu\nu} dx^{\mu}dx^{\nu}\,.
\end{multline}
Since the viscous corrections to $\tau_{\mu\nu}$ can come about only through the tensor modes $\delta \pi^{(2)}_{\mu\nu}$, it is sufficient to compute $\tau_{23}$. Using \eqref{E:gmnToTmn}, we find
\begin{equation}
\label{E:tauval}
	\kappa_5^2 \tau_{23} = -\frac{1}{2 b^3} \sigma_{23} + 2 \lim_{r \to \infty} r^4 \delta \pi_{23}^{(2)}
		+ \frac{1}{2 q^2} \lim_{r \to \infty} r^3 \left( f^{(2)} -  \delta_{\Delta\,3/2} s^{(2)}\right) \sigma_{23} + \frac{1}{2 q^2} \lim_{r \to \infty} r \gamma \sigma_{23}^{s}  +\pdt
\end{equation}
Here $\delta_{\Delta\,3/2}$ is the Kronecker delta function. It appears because $\lim_{r \to \infty} r^3 s^{(2)}$ is non vanishing only for $\Delta=3/2$.

Since the equation of motion for $\delta\pi^{(2)}_{\mu\nu}$ decouples from the other components of the metric we will focus exclusively on it. It is convenient to define
\begin{equation}
\label{E:decomposition}
	\delta\pi^{(2)}_{\mu\nu} = \delta\pi_n(r) \sigma^n_{\mu\nu} + \delta\pi_s(r) \sigma^s_{\mu\nu}
\end{equation}
where $\sigma_{\mu\nu}^n$ and $\sigma_{\mu\nu}^s$ were given in \eqref{E:sigmas}.
Using \eqref{E:decomposition} the equation of motion for $\delta\pi^{(2)}_{\mu\nu}$ can be decomposed in  two,\footnote{%
Equation \eqref{E:omeganEOM} can be read off the $q^{-2}$ term in an appropriate large $q$ series expansion of equation (27) in  \cite{Haack:2008xx}.
}
\begin{align}
\label{E:omegasEOM}
	\frac{d}{dr} \left( r(b^4 r^4- 1)) \frac{d }{dr} \delta\pi_s\right) &= - b^4 \frac{d}{dr} \left(\gamma r\right) \\
\label{E:omeganEOM}
	\frac{d}{dr} \left((b^4 r^4 - 1)  \frac{d}{dr} \delta\pi_n \right) &= -\frac{d}{dr} \left( b^4 r^4 \left( f^{(2)} - (1-(r b)^{-4})s^{(2)} \right) \frac{d}{dr} \delta \pi^{(0)} \right)\,.
\end{align}

Integrating \eqref{E:omeganEOM} once and requiring that $\delta \pi_n$ is well defined at the horizon, we find that
\begin{equation}
\label{E:pinsolution}
	\lim_{r \to \infty} r^4 \delta\pi_n = -\frac{1}{4} \lim_{r \to \infty} r^3 \left( f^{(2)} - \delta_{\Delta\,3/2} s^{(2)}\right) + \frac{3}{16} \frac{f^{(2)}(1/b)}{b^3} + \pdt\,.
\end{equation}
Inserting \eqref{E:pinsolution} into \eqref{E:tauval} and comparing to \eqref{E:ViscousCorrections} we conclude that
\begin{equation}
\label{E:etaVal}
	2 \kappa_5^2 \eta = \frac{1}{b^3} - \frac{3 f^{(2)}(1/b)}{4 b^3 q^2}\,.
\end{equation}
The $\mathcal{O}(q^{-2})$ corrections to the shear viscosity given in \eqref{E:etaVal} are expected. Once $f^{(2)} \neq 0$ the location of the horizon is shifted to 
\begin{equation}
\label{E:newhorizon}
	r = \frac{1}{b} - \frac{f^{(2)}(1/b)}{4 b q^2}\,.
\end{equation}
Inserting \eqref{E:newhorizon} into the Bekenstein-Hawking formula for the entropy density, one finds that the $q^{-2}$ corrections to the shear viscosity in \eqref{E:etaVal} exactly cancel the $q^{-2}$ corrections to the entropy density coming from the shift in the location of the horizon \eqref{E:newhorizon} so that the ratio of the shear viscosity to entropy density retains its universal value \eqref{E:Universal} up to order $\mathcal{O}(q^{-3})$.

To obtain $\eta_s$ 
we integrate \eqref{E:omegasEOM} once. Requiring that $\delta \pi_s$ is finite at $r b = 1$ together with \eqref{E:gammahzero} implies 
\begin{equation}
\label{E:pissolution}
	\lim_{r \to \infty} r^4 \delta\pi_s = -\frac{1}{4} \lim_{r \to \infty} r \gamma\,.
\end{equation}
Inserting \eqref{E:pissolution} into \eqref{E:tauval} and comparing to \eqref{E:ViscousCorrections} we obtain
\begin{equation}
\label{E:Zeroetas}
	\eta_s = 0\,.
\end{equation}

\subsection{An explicit solution}
\label{S:ExplicitO1}
We now turn to solving \eqref{E:deltaGEOM} and \eqref{E:deltarhoEOM} near the phase transition for the special case $m^2=-4$ discussed in \cite{Herzog:2010vz} and reviewed in section \ref{S:Explicit}. As in section \ref{S:Explicit} we formally expand $\delta G$ and $\delta \rho$ around $\mu = \mu_0 = 2/b$,
\begin{align}
\label{E:deltaExpansion}
\begin{split}
		\delta G &= \sum_{\overline{n}=0}^{\infty} \delta G_0^{\oss{2n}}(r b) (\mu b - 2)^n \\
		\delta \rho &= \sum_{\overline{n}=0}^{\infty} \delta \rho^{\oss{2n-1}}(r b) (\mu b - 2)^{(2n-1)/2}\,.
\end{split}
\end{align}
We will soon see that $\delta \rho^{\oss{-1}} \neq 0$ which implies that $\delta\rho$ diverges close to the phase transition. Thus, our analysis is valid only for when the gradient corrections to the various fields are smaller than the relative magnitude of the chemical potential, $\partial_{\mu} u^{\mu} \ll (\mu b - 2)$.
Since the equations of motion for the delta'd quantities  \eqref{E:deltaGEOM} and \eqref{E:deltarhoEOM} are linearized versions of their non delta'd counterparts \eqref{E:ProbeEOM}, inserting the expansion \eqref{E:deltaExpansion} into \eqref{E:ProbeEOM} yields a set of equations very similar to those obtained for $G_0$ and $\rho$ which were described in section \ref{S:Explicit}. 
The solution for the leading order terms in \eqref{E:deltaExpansion} is given by
\begin{align}
\begin{split}
\label{E:deltasollead}
		\delta \rho^{\oss{-1}}(x) &= +\frac{9 \sqrt{3}}{49 (1+x^2)}\partial_{\mu}N^{\mu} b^2 \\
		\delta G^{\oss{0}}(x) & = -\frac{2 x}{3(1+x^2)}\partial_{\nu}u^{\nu} - \frac{13 x^4+13 x^2 - 54}{49 x^2(1+x^2)} \partial_{\mu}N^{\mu} b \ .
\end{split}
\end{align}
We also need the solutions for $\delta\rho^{\oss{1}}$ and $\delta G_0^{\oss{2}}$, but they are too lengthy to repeat here.

To solve the equations of motion for the vector modes we insert the expansion
\begin{equation}
	\delta g_{\mu} = \sum_{n=0}^{\infty} \delta g_{\mu}^{\oss{2n}}(r b) \left(\mu b - 2\right)^{n}
\end{equation}
into \eqref{E:VectorModesO1} and collect terms with equal powers of $\mu b -2$. The first couple of regular solutions which vanish at the asymptotically AdS boundary are
\begin{align}
\begin{split}
\label{E:deltaVsollead}
	\delta g_{\mu}^{\oss{0}}(x)  & = \frac{1}{4\pi} \left(\pi - 2 \arctan x + 2\ln(1+x) - \ln(1+x^2) \right) P_{\mu}^{\phantom{\mu}\nu} \partial_{\nu} \frac{\mu}{T} - \frac{2 x}{1+x^2} u^{\alpha} \partial_{\alpha} u_{\mu} \\
	\delta g_{\mu}^{\oss{2}}(x) & =  -\frac{3}{2\pi (1+x^2)} \left(\pi - 2 \arctan x + 2\ln(1+x) - \ln(1+x^2) -2\right) P_{\mu}^{\phantom{\mu}\nu} \partial_{\nu} \frac{\mu}{T} - \frac{x (x^2-5)}{(1+x^2)^2} u^{\alpha} \partial_{\alpha} u_{\mu} \,.
\end{split}
\end{align}
The expression for $\delta g_{\mu}^\oss{0}$ has the expected form discussed in section \ref{S:Vsector} which ensures that $\kappa$ is continuous across the phase transition.

With the solutions $\delta \rho^{\oss{-1}}$, $\delta G^{\oss{0}}$, $\delta\rho^{\oss{1}}$, and $\delta G_0^{\oss{2}}$
in hand we can compute the correction to the Josephson condition
\begin{equation}
	u^{\mu} \partial_{\mu} \phi = - \mu - \lim_{r\to\infty} \delta G = -\mu +\left(  \frac{13}{49} + 
	\frac{1823-3004 \ln 2}{343} (\mu b -2) \right)
	b \partial_{\mu}N^{\mu}\,.
\end{equation}
Comparing this to the third line of \eqref{E:ViscousCorrections} and using
\eqref{E:rhoandrhos} we obtain \eqref{E:FinalZeta3}. As expected, $\zeta_3$ has a power law behavior close to the phase transition. 
Using
\begin{equation}
\label{E:OVEV}
	\kappa_5^2 |\langle O_{\psi} \rangle|  = C_{\Delta} \lim_{r \to \infty} r^{\Delta} \left(\rho + \delta\rho \right) + \mathcal{O}(\partial^2) 
\end{equation}
we find that the viscous corrections to $|\langle O_{\psi} \rangle |$, call them $\delta |\langle O_{\psi} \rangle |$, take the form
\begin{equation}
	\kappa_5^2 \delta |\langle O_{\psi} \rangle | =
	-\frac{9 \sqrt{3}C_{\Delta} \partial_{\nu} N^{\nu}}{49 \sqrt{\left(\mu b-2\right)}} \left( 1 - \frac{70993 - 116496  \ln 2}{3024} \left(\mu b-2\right) + \mathcal{O} \left( \left(\mu b-2\right)^{2}\right) \right)\,.
\end{equation}
Finally, using \eqref{E:deltaVsollead} together with \eqref{E:JVEV} and \eqref{E:ViscousCorrections} we compute
\begin{equation}
	\kappa_5^2 \kappa = 	\frac{1}{2 \pi b^2} - \frac{3}{\pi b^2}(\mu b - 2) + \mathcal{O}\left( (\mu b - 2)^2 \right)
\end{equation}
which gives us \eqref{E:FinalKappa}.

The leading order expressions for $\kappa$ and $\zeta_3$ can be obtained from the Kubo formulas \eqref{Kubozeta3} and \eqref{KubokappaT} and the analysis of \cite{Herzog:2010vz}. We present this analysis in appendix \ref{A:Kubo}.

\section{Discussion}
In this work we have constructed a slowly varying, spacetime dependent, holographic superfluid and computed its transport coefficients. The resulting stress-energy tensor and charged current of the gauge-theory superfluid fits nicely with the relativistic version of the Tisza-Landau two fluid model. Apart from the shear viscosity and diffusion coefficient, the conformally invariant two-fluid model allows for another transport coefficient which we denoted by $\zeta_3$. In section \ref{S:Sound} we saw that $\zeta_3$, together with $\kappa$ and $\eta$, is responsible for the attenuation of second sound.  Another physical role of $\zeta_3$ can be understood by considering entropy production. With the assumptions specified in \ref{S:Viscous}, the divergence of the entropy current $J_{s}^{\mu}$ can be read off of \eqref{E:viscousudT}. It is given by
\begin{equation}
\label{E:EntropyProduction}
	\partial_{\mu}J_{s}^{\mu} = \frac{\zeta_3}{T} \left( \partial_{\mu} \left( \rhos n^{\mu} \right)\right)^2 + \kappa P^{\mu\nu} \left(\partial_{\mu} \frac{\mu}{T} \right) \left( \partial_{\nu} \frac{\mu}{T}  \right)+ \frac{\eta}{T} \sigma^n_{\mu\nu}\partial^{\mu}u^{\nu}\,.
\end{equation}
Thus, for $n^{\mu} \ll 1$, $\zeta_3$ is responsible for entropy production due to compressibility of the superfluid component. Indeed, expression \eqref{E:EntropyProduction} can be compared with the divergence of the entropy current in a normal (non superfluid) relativistic fluid which is not conformally invariant, c.f. \cite{Landau},
\begin{equation}
	\partial_{\mu} J_{s}^{\mu} =  \frac{\zeta_1}{T} \left( \partial_{\mu} u^{\mu} \right)^2 + \kappa P^{\mu\nu} \left( \partial_{\mu} \frac{\mu}{T}\right) \left( \partial_{\nu} \frac{\mu}{T} \right) + \frac{\eta}{T} \sigma^n_{\mu\nu}\partial^{\mu}u^{\nu}
\end{equation}
where $\zeta_1$, the bulk viscosity, plays a role very similar to $\zeta_3$. 

The power-law divergence of $\zeta_3$ near the phase transition that we observed in our holographic computation does not imply that entropy production diverges there. While $\zeta_3 \sim \mathcal{O}\left((\mu b - \mu_0 b)^{-1}\right)$, the superfluid density vanishes, $\rhos \sim \mathcal{O}\left( \mu b - \mu_0 b \right)$. The extra power of $\rhos$ in \eqref{E:EntropyProduction} guarantees that $\partial_{\mu}J^{\mu}_{s}$ vanishes close to the phase transition.\footnote{We thank D.~Son for pointing this out to us.} A similar observation can be made regarding the dependence of $\zeta_3$ on $\kappa_5^2$. The boundary value of $G_{\mu}$ is independent of Newton's constant. Since $\zeta_3$ appears together with $\rhos$ then $\rhos \sim \kappa_5^{-2}$ (which follows from \eqref{E:JVEV}) implies that $\zeta_3 \sim \kappa_5^2$.

We note that too close to the phase transition one expects the hydrodynamic approximation to break down since the Landau-Ginzburg potential becomes approximatety flat. Our expansion near $T_0$ is then valid only when the hydrodynamic fluctuations are smaller then the relative chemical potential. This fact also came into play when we discussed the expansion of the bulk scalar near $T_0$ in section \ref{S:ExplicitO1}.

The analysis carried out in this work is valid for configurations where the charge of the scalar field is large. From the structure of the perturbative expansion, it seems likely that $\zeta_3$ will  diverge near $\mu_0$ for smaller values of $q$. As long as the boundary value of $\delta G_{\mu}$ does not vanish close to the phase transition, the fact that $\rhos$ vanishes there implies through \eqref{E:Josephson} that $\zeta_3$ must diverge. We have no argument for having $\eta_s=0$ in the fully backreacting case. We feel that the result \eqref{E:Zeroetas} valid at order $\mathcal{O}(q^{-2})$ together with \eqref{E:viscousudT} make it unlikely that $\eta_s$ will be $q$ dependent.

\label{S:Discussion}

\section*{Acknowledgments}
We thank M.~Ammon and D.~Son for useful discussions and M.~Haack and S.~Minwalla for correspondence. PS and AY thank the organizers of the workshop ``AdS Holography and the Quark-Gluon Plasma" at the ESI in Vienna for hospitality and partial support. PS thanks Princeton University for hospitality.  CH is supported in part by the US NSF under Grants No. PHY-0844827 and PHY-0756966. CH thanks the Sloan Foundation for partial support. PS is supported in part by the Belgian Federal Science Policy Office through the Interuniversity Attraction Pole IAP VI/11 and by FWO-Vlaanderen through project G011410N. AY is supported in part by the Department of Energy under Grant No. DE-FG02-91ER40671.

\begin{appendix}
\section{Bulk to boundary mapping}
\label{A:BtoB}
The prescription for mapping bulk fields to boundary expectation values can be found in sections \ref{S:AdSSS} to \ref{S:backreaction}. In what follows we collect these results for ease of access.
The energy momentum tensor of the boundary theory is given by
\begin{multline}
\tag{\ref{E:gmnToTmn}}
	\kappa_5^2 \langle T_{\mu\nu} \rangle  =  -\lim_{r \to \infty} r^2 \Bigg( K_{\mu\nu} - K \gamma_{\mu\nu} + 3 \gamma_{\mu\nu} 
	- \frac{1}{2} \gamma_{\mu\nu} m_{\Delta} |\psi|^2 
\\
	- \frac{1}{4} \gamma_{\mu\nu} \tilde{m}_{\Delta} \left(\psi^* n^{\alpha}\partial_{\alpha} \psi  - \left(\psi^*n_{\mu}\partial_{\nu} \psi + \psi^*n_{\nu}\partial_{\mu} \psi\right)   + \hbox{c.c.} \right)
\Bigg) \,.
\end{multline}
The norm of the scalar follows from \eqref{E:psiToO},
\begin{equation}
\tag{\ref{E:OVEV}}
	\kappa_5^2 |\langle O_{\psi} \rangle|  = C_{\Delta} \lim_{r \to \infty} r^{\Delta} \left(\rho + \delta\rho \right) + \mathcal{O}(\partial^2) \,.
\end{equation}
The expectation value of the gradient of the Goldstone boson is given by
\begin{equation}
\tag{\ref{E:JosephsonCorrection}}
	u^{\mu} \partial_{\mu} \phi = - \lim_{r \to \infty} \left( G_0 + \delta G \right) \,.
\end{equation}
The expectation value of the current follows from \eqref{E:AToJ}
\begin{equation}
\tag{\ref{E:JVEV}}
	\kappa_5^2 \langle J_{\mu} \rangle  = \lim_{r \to \infty} r^2 \left( G_{\mu} + \delta G_{\mu} \right)
	+ \pdt\,.
\end{equation}
Using our definition of the charge density in section \ref{S:Superfluids} we find that 
\begin{align}
\tag{\ref{E:rhoandrhos}}
\begin{split}
	\kappa_5^2 \rhot &= -\lim_{r \to \infty}  r^2 G_0 +\pdt \\
	\kappa_5^2 \mu^{-1} \rhos & = -\lim_{r \to \infty} r^2 g +\pdt \,.
\end{split}
\end{align}

\section{Two Point Functions from Gravity}
\label{A:Kubo}

In this appendix we work out the leading term for $\kappa$ and $\zeta_3$ using the Kubo relations in section \ref{S:Kubo}. Most of the Greens function needed for carrying out this computation have already been computed in \cite{Herzog:2010vz}. We briefly go over the notation of \cite{Herzog:2010vz} and compute the remaining Greens function.

We start with the action (\ref{E:action}) and 
work in a weak gravity (or probe) limit $q \gg 1$ in which gravity decouples from the scalar and gauge fields (or Abelian-Higgs sector) \cite{Hartnoll:2008vx}.  
In this limit, a solution to Einstein's equations is a black brane:
\be
ds^2 = \frac{L^2}{u^2} \left( -f(u) dt^2 + dx^2 + dy^2 + dz^2 + \frac{du^2}{f(u)} \right) \ ,
\label{metric}
\ee
where $f(u) = 1-(u/u_h)^4$. At $u=u_h>0$, there is a horizon, while in the limit $u \to 0$, the space asymptotically becomes anti-de Sitter. Since we will use linear response to compute the transport coefficients it is sufficient to work in the Fefferman-Graham coordinate system. In \eqref{E:MetricAnsatz} we used an ingoing Edington-Finckelstein coordinate system.
For convenience, we will set $2 \kappa_5^2 = L =u_h = q = 1$ in what follows. We also work in a gauge where $A_5=0$ and consistently set $\psi$ to be real.

When $m^2 = -4$ the near boundary expansion for $\psi$ and $A_{\mu}$ takes the form
\begin{align}
\psi &=   \left(  \psi^{(b)} u^2 \ln(u/\lambda) - \langle O_\psi \rangle u^2 + \ldots \right)
\label{smallPsi} \ , \\
A_\mu &= A_\mu^{(b)} + \frac{1}{2} \langle J_\mu \rangle u^2 + \ldots \ , 
\label{smallAmu}
\end{align}
where we have introduced a UV cutoff $\lambda \ll 1$.
In this coordinate system, in place of \eqref{E:OVEV} and \eqref{E:JVEV} we have
\be
\langle J^\mu \rangle = \lim_{u \to 0} \frac{1}{\sqrt{-g^{(b)}}} \frac{\delta S}{\delta A^{(b)}_\mu} \ , \; \; \;
\myop = \lim_{u \to 0} \frac{1}{\sqrt{-g^{(b)}}} \frac{\delta S}{\delta {\psi^{(b)}}^*} \ ,
\label{onepoint}
\ee
where
$
g_{\mu\nu}^{(b)} = \lim_{u \to 0} u^2 g_{\mu\nu}
$
is the Minkowski metric.
Since we are working in a gauge where $\psi$ is real, we can identify $A_t^{(b)}$ with the chemical potential $\mu$ and $A_i^{(b)}$ with a superfluid velocity $\partial_i\phi$.\footnote{%
There is a minus sign discrepancy between this identification and the one in \cite{Herzog:2008he}.  Here we define $A_i(0) \equiv \xi_i$ whereas in \cite{Herzog:2008he}, $\xi_i$ was defined to be the phase gradient of the scalar.}

As described in section \ref{S:Explicit} we can solve for $\psi$ and $A_{\mu}$ near the phase transition $\mu/T= 2\pi$. Following the notation of \cite{Herzog:2010vz} instead of using $\mu/T-2\pi$ as a small parameter, we use
\be
\epsilon \equiv -\sqrt{2} \langle O_\psi \rangle \ .
\ee
At equilibrium in the absence of a superfluid velocity, we have the near boundary expansions
\begin{align}
\sqrt{2} \psi =& \epsilon u^2 + \mathcal{O}(u^4) \\
A_t =&
\left( 2 + \delta \mu_2 \epsilon^2 + \delta \mu_4 \, \epsilon^4 + \mathcal{O}(\epsilon^6) \right) 
-
\left(2 +\left(\frac{1}{8} + \delta \mu_2 \right) \epsilon^2 + \right. \\
& \left. \left(\frac{-5  +6\log 2}{1152} + \delta \mu_4 \right) \epsilon^4
+ O(\epsilon^6) \right) u^2
+ O(u^4) \ , \nonumber
\end{align}
where
\be
\delta \mu_2 = \frac{1}{48} \ , \; \; \;
\delta \mu_4 = 
\left(\frac{253}{55{,}296} + \frac{7  \log 2}{1152} \right) \ .
\ee

To compute the Greens functions we consider the fluctuations of the fields:
 \begin{align}
 \begin{split}
\delta A_t(u,t,x) =& a_t(u) e^{-i \omega t + i k x}  \\
\delta A_x(u,t,x) =& a_x(u) e^{-i \omega t + i k x}  \\
\delta \psi(u,t,x) =& \psi(u) e^{-i \omega t + i k x} / \sqrt{2} \ .
\end{split}
\end{align}
The Greens functions can be computed
from the coefficient of the $u^2$ term in a near boundary expansion of the fluctuations (see \cite{Herzog:2010vz} for more details). We find:
\begin{align}
\begin{split}
\label{E:Greens}
\frac{G^{tt}}{k^2} = \frac{G^{tx}}{\omega k} = \frac{G^{xx}}{\omega^2} =&
- \frac{2}{{\mathcal P}}
\left[ 
48 i k^4 + 7 i  \epsilon^4 + 148  \epsilon^2 \omega - 480 i \omega^2 
+ 8 k^2 (5 i  \epsilon^2 + 12 \omega)
\right]
\\
G^{Ot} =G^{tO} =&  
\frac{8 i \epsilon }{{\mathcal P}} \left[ 12 k^4 + (7 \epsilon^2 - 120 i \omega) \omega
+ 3 k^2 ( \epsilon^2 + 16 (1-i) \omega) \right] 
\\
G^{ O x} =G^{x O} =& + 
\frac{i k \epsilon }{ {\mathcal P}} \left[
48 k^4 + \epsilon^4 + 12 (2-3i)\epsilon^2 \omega - 96(3 + 4i) \omega^2 + \right.  
\\
& \hspace{50mm}\left. + 16 k^2 (\epsilon^2 + 6(1-2i) \omega 
\right] 
\\
G^{OO} =& -\frac{2 i }{{\mathcal P}}
\left[ 96 k^4 + \epsilon^4 + 16(3-2i) \epsilon^2 \omega - 192 (1 + 3i) \omega^2 + \right.
\\
& \hspace{50mm} \left. 4 k^2 ( 7 \epsilon^2 + 72 (1-i) \omega) \right] 
\\
G^{O \overline O} = G^{\overline O O} =& \frac{2 i \epsilon^2}{{\mathcal P}} (4 k^2 + \epsilon^2 - 32 i \omega) 
\\
\end{split}
\end{align}
where the pole structure is given by 
\begin{equation}
\label{E:Pole}
{\mathcal P} =
960 \omega^3 + 56i (12 k^2 +  \epsilon^2) \omega^2 
-12 (16 k^2 + 3  \epsilon^2) k^2 \omega 
- i (48 k^4 + 16 k^2  \epsilon^2 +  \epsilon^4 ) k^2 \ .
\end{equation}
Expressions \eqref{E:Greens} and \eqref{E:Pole} are the leading order results in a limit where 
$\epsilon^2 \sim k^2 \sim \omega \ll 1$ are all small and of the same order.
The complex conjugate Green's functions involving $\overline O_\psi$ 
can be obtained by sending $\omega \to -\omega$, $k\to -k$, and $i \to -i$.  
These Green's functions obey the Ward identities $-\omega G^{tt} + k G^{tx} = 0$, $-\omega G^{tx} + k G^{xx} = 0$, $- \omega G^{O t} + k G^{O x} = \langle O_\psi \rangle$, and $- \omega G^{\overline O t} + k G^{\overline O x} = -\langle O_\psi \rangle$. 
From the two-point functions involving $O$ and $\overline O$, we can construct a two-point function for the phase of the condensate:
\begin{align}
\begin{split}
G^{\phi \phi} =& 
\frac{1}{2\epsilon^2} \left( G^{O O} + G^{\overline O \overline O} - G^{\overline O O} - G^{O \overline O} \right)
\\
=&
-\frac{4 i}{\epsilon^2 {\mathcal P}}(48 k^4 + \epsilon^4 + 16 k^2 (\epsilon^2 - 9 i \omega) - 32 i \epsilon^2 \omega - 96 \omega^2) \ .
\end{split}
\end{align}

By direct computation, we find the following Kubo relations
\begin{subequations}
\begin{align}
\lim_{\omega \to 0} \lim_{k \to 0} \frac{\omega^2}{k^2} \operatorname{Re} G^{tt} =\lim_{\omega \to 0} \lim_{k \to 0}  \operatorname{Re} G^{xx} =& -\frac{\epsilon^2}{4} \ , 
\label{GxxKuboRe} 
\\
\lim_{\omega \to 0} \lim_{k \to 0} \frac{\omega}{k^2} \operatorname{Im}  G^{tt} =\lim_{\omega \to 0} \lim_{k \to 0} \frac{1}{k}  \operatorname{Im}  G^{xt} =& 1 \ , 
\label{GxxKuboIm}
\\
\lim_{\omega \to 0} \lim_{k \to 0} \omega^2 \operatorname{Re} \, G^{\phi \phi} =& - \frac{1}{14} \ , 
\label{GphiKuboRe}
\\
\lim_{\omega \to 0} \lim_{k \to 0} \omega \operatorname{Im} \, G^{\phi \phi} =& \frac{52}{49} \frac{1}{ \epsilon^2} \ .
\label{GphiKuboIm}
\end{align}
\end{subequations}
Comparing with (\ref{Kuborhosmu}), (\ref{KubokappaT}), (\ref{Kubomurho}), and (\ref{Kubozeta3}), the right hand side of these Kubo relations should correspond to $-\rhos/\mu$, $\kappa/T$, $-\partial \mu/ \partial \rhot$, and $\zeta_3$ respectively.  Restoring factors of $\kappa_5$ and $T$, we find that they do agree at leading order with (\ref{E:rhoandrhosseries}), (\ref{E:FinalKappa}) and (\ref{E:FinalZeta3}). 
Since (\ref{GxxKuboIm}) is independent of the value of $\rhos$, the relation implies that $\kappa$ is continuous through the phase transition.

Another way of extracting $\kappa$ and $\zeta_3$ is through attenuation of fourth sound.
The solutions of ${\mathcal P}=0$ give poles of the Green's functions from which we can infer the dispersion relations
\be
\omega = \pm \frac{k \epsilon}{2 \sqrt{14}} - \frac{33 i k^2}{196} + \ldots  \ , \; \; \;
 \omega  = - \frac{i k^2}{2} \ ,
\ee
for $T \lesssim T_0$ and $T \gtrsim T_0$ respectively.
These expressions should be compared with (\ref{fourthsounddispersion})   and (\ref{E:fourthsoundAlternate}).
Note that $\rhos = 0$ and $\partial \rhot / \partial \mu = 2$ for $T \gtrsim T_0$ while for $T \lesssim T_0$, these quantities are given by (\ref{GxxKuboRe}) and (\ref{GphiKuboRe}).
If we read off the value of $\kappa$ from the $T \gtrsim T_0$ result, we find agreement with (\ref{GxxKuboIm}).  If we then assume that $\kappa$ is continuous through the phase transition, we can deduce the value of $\zeta_3$ and find agreement with (\ref{GphiKuboIm}).

\section{Formulae for the Explicit Solution near $T_0$}
\label{A:exactsols}

This appendix provides the first few terms, in an expansion in $(\mu - \mu_0)b$, of the static background solution near $T_0$ described in section \ref{S:Explicit}.
For the scalar quantities $G_0$ and $\rho$ we have
\begin{align}
\label{E:Gsandrhos}
\begin{split}
	G_0^\oss{0}(x) =& 1-x^{-2} \\
	G_0^\oss{2}(x) =&  \frac{x^4 - 6 x^2 +5}{x^2(x^2+1)} \\
	G_0^\oss{4}(x) =&  -\frac{309 x^4 - 56 x^2 - 253}{4 x^2 (x^2+1)^2} + \frac{(72 x^2 - 96) \ln 2}{x^2 (x^2+1)}  + \frac{24 x^2 \ln(1+x^{-2})}{x^2+1} \\
	\rho^{\oss{1}}(x) =&  \frac{4 \sqrt{3}}{1+x^2} \\
	\rho^{\oss{3}}(x) =&\frac{253 x^2+157}{4\sqrt{3} (x^2+1)^2} - \frac{28 \sqrt{3} \ln 2}{x^2+1} + \frac{4 \sqrt{3} \ln \left(1+\frac{1}{x^2}\right)}{x^2+1}\,.
	\\
\end{split}
\end{align}
For $g$, the first few terms are
 \begin{align}
\begin{split}
\label{E:gs}
	g^{\oss{0}}(x) =& 1 \\
	g^{\oss{2}}(x) =& -\frac{6}{x^2+1}\\
	g^{\oss{4}}(x) =& 6\left(\text{Li}_2\left(-x^2\right)+\text{Li}_2 \left(1-x^2\right)+\text{Li}_2\left(\frac{x^2+1}{2} \right)\right)
	\\
	&-\frac{289 x^2+229}{4 \left(x^2+1\right)^2}-6 \ln (2)
   \ln \left(1-x^2\right)+\frac{12 x^2 \ln
   \left(1+x^{-2}\right)+84 \ln
   (2)}{x^2+1}+3 \ln ^2(2)
   \\
   &+3\left(2 \ln \left(x^2
   \left(1-x^2\right)\right)-\ln
   \left(x^2+1\right)\right) \ln
   \left(x^2+1\right) \ ,
   \\
\end{split}
\end{align}	
For $s^{(2)}$ and $f^{(2)}$, the first few terms are
\begin{align}
\begin{split}
	s^{(2,\overline{0})}(x) =& 0\\
	s^{(2,\overline{2})}(x) =& -\frac{8}{\left(1+x^2\right)^2} \\
	s^{(2,\overline{4})}(x) =& -\frac{48 x^6+421 x^4+618 x^2 +221}{3(1+x^2)^4} + \frac{112 \ln 2}{(1+x^2)^2} + \frac{16 x^2 (2+x^2) \ln(1+x^{-2})}{(1+x^2)^2} \\
	f^{(2,\overline{0})}(x) =& \frac{4 }{3 x^6}\\
	f^{(2,\overline{2})}(x) =& \frac{4 \left(-5 + 19 x^2 \right)}{3 x^6 (1+x^2)} \\
	f^{(2,\overline{4})}(x)  = &-\frac{4 \left(24 x^{10} + 60 x^8 -208 x^6 - 335 x^4 - 38 x^2 + 57 \right)}{3 x^6 (1+x^2)^3} \\
		& - \frac{ 64 (5x^2-2) \ln 2}{x^6(x^2+1)} + \frac{32 (x^4+x^2-2) \ln(1+x^{-2})}{x^2 (x^2+1)}  \,.
\end{split}
\end{align}
Finally, the vector mode corrections to the metric are given by
\begin{align}
\begin{split}
	\gamma^\oss{0}(x) =& 0\\
	\gamma^\oss{2}(x) =& \frac{6(x^2-1)}{x^4(x^2+1)}\\
	\gamma^{\oss{4}}(x)=& -\frac{x^4-1}{x^4} g^{\oss{4}} + \frac{x^2-1}{2 x^2} G_0^{\oss{4}} - \frac{x^2-1}{8(x^2+1)^2 x^4} \left(413 + 538 x^2 + 269 x^4\right) + \frac{x^2-1}{x^4} 60 \ln 2 
	\ .
\end{split}
\end{align}
Note that the leading order corrections to the backreacted metric $f^{(2,\overline{0})}$and $s^{(2,\overline{0})}$ together with the leading order expression \eqref{E:MetricAnsatz} generate an exact solution to the Einstein equations---the Reissner-Nordstrom black hole. This solution describes the fully backreacted geometry at high temperatures when $\psi$ is not condensed.

\end{appendix}

\bibliographystyle{ssg}
\bibliography{SFBV}

\begingroup\raggedright\begin{thebibliography}{10}

\bibitem{Maldacena:1997re}
J.~M. Maldacena, ``{The Large N limit of superconformal field theories and
  supergravity},'' {\em Adv.Theor.Math.Phys.} {\bf 2} (1998) 231--252,
  \href{http://xxx.lanl.gov/abs/hep-th/9711200}{{\tt hep-th/9711200}}.

\bibitem{Gubser:1998bc}
S.~Gubser, I.~R. Klebanov, and A.~M. Polyakov, ``{Gauge theory correlators from
  noncritical string theory},'' {\em Phys.Lett.} {\bf B428} (1998) 105--114,
  \href{http://xxx.lanl.gov/abs/hep-th/9802109}{{\tt hep-th/9802109}}.

\bibitem{Witten:1998qj}
E.~Witten, ``{Anti-de Sitter space and holography},'' {\em
  Adv.Theor.Math.Phys.} {\bf 2} (1998) 253--291,
  \href{http://xxx.lanl.gov/abs/hep-th/9802150}{{\tt hep-th/9802150}}.

\bibitem{Bhattacharyya:2008jc}
S.~Bhattacharyya, V.~E. Hubeny, S.~Minwalla, and M.~Rangamani, ``{Nonlinear
  Fluid Dynamics from Gravity},'' {\em JHEP} {\bf 0802} (2008) 045,
  \href{http://xxx.lanl.gov/abs/0712.2456}{{\tt 0712.2456}}.

\bibitem{Hartnoll:2008vx}
S.~A. Hartnoll, C.~P. Herzog, and G.~T. Horowitz, ``{Building a Holographic
  Superconductor},'' {\em Phys.Rev.Lett.} {\bf 101} (2008) 031601,
  \href{http://xxx.lanl.gov/abs/0803.3295}{{\tt 0803.3295}}.

\bibitem{Herzog:2008he}
C.~P. Herzog, P.~K. Kovtun, and D.~T. Son, ``{Holographic model of
  superfluidity},'' {\em Phys. Rev.} {\bf D79} (2009) 066002,
  \href{http://xxx.lanl.gov/abs/0809.4870}{{\tt 0809.4870}}.

\bibitem{Basu:2008st}
P.~Basu, A.~Mukherjee, and H.-H. Shieh, ``{Supercurrent: Vector Hair for an AdS
  Black Hole},'' {\em Phys.Rev.} {\bf D79} (2009) 045010,
  \href{http://xxx.lanl.gov/abs/0809.4494}{{\tt 0809.4494}}.

\bibitem{MembraneParadigm}
{Thorne, K.~S., Price, R.~H., \& MacDonald, D.~A.}, ed., {\em {Black holes: The
  membrane paradigm}}.
\newblock {Yale University Press}, {New Haven, CT.}, 1986.

\bibitem{Policastro:2001yc}
G.~Policastro, D.~Son, and A.~Starinets, ``{The Shear viscosity of strongly
  coupled N=4 supersymmetric Yang-Mills plasma},'' {\em Phys.Rev.Lett.} {\bf
  87} (2001) 081601, \href{http://xxx.lanl.gov/abs/hep-th/0104066}{{\tt
  hep-th/0104066}}.

\bibitem{Kovtun:2004de}
P.~Kovtun, D.~Son, and A.~Starinets, ``{Viscosity in strongly interacting
  quantum field theories from black hole physics},'' {\em Phys.Rev.Lett.} {\bf
  94} (2005) 111601, \href{http://xxx.lanl.gov/abs/hep-th/0405231}{{\tt
  hep-th/0405231}}.

\bibitem{Buchel:2003tz}
A.~Buchel and J.~T. Liu, ``{Universality of the shear viscosity in
  supergravity},'' {\em Phys.Rev.Lett.} {\bf 93} (2004) 090602,
  \href{http://xxx.lanl.gov/abs/hep-th/0311175}{{\tt hep-th/0311175}}.

\bibitem{Buchel:2004qq}
A.~Buchel, ``{On universality of stress-energy tensor correlation functions in
  supergravity},'' {\em Phys.Lett.} {\bf B609} (2005) 392--401,
  \href{http://xxx.lanl.gov/abs/hep-th/0408095}{{\tt hep-th/0408095}}.

\bibitem{Brustein:2009rn}
R.~Brustein and A.~Medved, ``{Proof of a universal lower bound on the shear
  viscosity to entropy density ratio},'' {\em Phys.Lett.} {\bf B691} (2010)
  87--90, \href{http://xxx.lanl.gov/abs/0908.1473}{{\tt 0908.1473}}.

\bibitem{Erdmenger:2010xm}
J.~Erdmenger, P.~Kerner, and H.~Zeller, ``{Non-universal shear viscosity from
  Einstein gravity},'' \href{http://xxx.lanl.gov/abs/1011.5912}{{\tt
  1011.5912}}.

\bibitem{Erdmenger:2008rm}
J.~Erdmenger, M.~Haack, M.~Kaminski, and A.~Yarom, ``{Fluid dynamics of
  R-charged black holes},'' {\em JHEP} {\bf 0901} (2009) 055,
  \href{http://xxx.lanl.gov/abs/0809.2488}{{\tt 0809.2488}}.

\bibitem{Banerjee:2008th}
N.~Banerjee, J.~Bhattacharya, S.~Bhattacharyya, S.~Dutta, R.~Loganayagam, {\em
  et.~al.}, ``{Hydrodynamics from charged black branes},''
  \href{http://xxx.lanl.gov/abs/0809.2596}{{\tt 0809.2596}}.

\bibitem{Torabian:2009qk}
M.~Torabian and H.-U. Yee, ``{Holographic nonlinear hydrodynamics from AdS/CFT
  with multiple/non-Abelian symmetries},'' {\em JHEP} {\bf 0908} (2009) 020,
  \href{http://xxx.lanl.gov/abs/0903.4894}{{\tt 0903.4894}}.

\bibitem{Son:2009tf}
D.~T. Son and P.~Surowka, ``{Hydrodynamics with Triangle Anomalies},'' {\em
  Phys.Rev.Lett.} {\bf 103} (2009) 191601,
  \href{http://xxx.lanl.gov/abs/0906.5044}{{\tt 0906.5044}}.

\bibitem{Eling:2010hu}
C.~Eling, Y.~Neiman, and Y.~Oz, ``{Holographic Non-Abelian Charged
  Hydrodynamics from the Dynamics of Null Horizons},''
  \href{http://xxx.lanl.gov/abs/1010.1290}{{\tt 1010.1290}}.

\bibitem{Haack:2008xx}
M.~Haack and A.~Yarom, ``{Universality of second order transport coefficients
  from the gauge-string duality},'' {\em Nucl.Phys.} {\bf B813} (2009)
  140--155, \href{http://xxx.lanl.gov/abs/0811.1794}{{\tt 0811.1794}}.

\bibitem{Tisza}
L.~Tisza, ``{Transport phenomena in helium {II}},'' {\em Nature} {\bf 141}
  (1938) 913.

\bibitem{Landau}
L.~D. Landau, ``{The theory of superfluidity of helium II},'' {\em J. Phys.
  USSR} {\bf 5} (1941) 71.

\bibitem{Khalatnikov198270}
I.~M. Khalatnikov and V.~V. Lebedev, ``Relativistic hydrodynamics of a
  superfluid liquid,'' {\em Physics Letters A} {\bf 91} (1982), no.~2 70 -- 72.

\bibitem{PhysRevD.45.4536}
B.~Carter and I.~M. Khalatnikov, ``Equivalence of convective and potential
  variational derivations of covariant superfluid dynamics,'' {\em Phys. Rev.
  D} {\bf 45} (Jun, 1992) 4536--4544.

\bibitem{Carter1992243}
B.~Carter and I.~M. Khalatnikov, ``Momentum, vorticity, and helicity in
  covariant superfluid dynamics,'' {\em Annals of Physics} {\bf 219} (1992),
  no.~2 243 -- 265.

\bibitem{Israel198179}
W.~Israel, ``Covariant superfluid mechanics,'' {\em Physics Letters A} {\bf 86}
  (1981), no.~2 79 -- 81.

\bibitem{Israel198277}
W.~Israel, ``Equivalence of two theories of relativistic superfluid
  mechanics,'' {\em Physics Letters A} {\bf 92} (1982), no.~2 77 -- 78.

\bibitem{Son:2000ht}
D.~T. Son, ``{Hydrodynamics of relativisic systems with broken continuous
  symmetries},'' {\em Int. J. Mod. Phys.} {\bf A16S1C} (2001) 1284--1286,
  \href{http://xxx.lanl.gov/abs/hep-ph/0011246}{{\tt hep-ph/0011246}}.

\bibitem{Pujol:2002na}
C.~Pujol and D.~Davesne, ``{Relativistic dissipative hydrodynamics with
  spontaneous symmetry breaking},'' {\em Phys. Rev.} {\bf C67} (2003) 014901,
  \href{http://xxx.lanl.gov/abs/hep-ph/0204355}{{\tt hep-ph/0204355}}.

\bibitem{Valle:2007xx}
M.~A. Valle, ``{Hydrodynamic fluctuations in relativistic superfluids},'' {\em
  Phys. Rev.} {\bf D77} (2008) 025004,
  \href{http://xxx.lanl.gov/abs/0707.2665}{{\tt 0707.2665}}.

\bibitem{Gusakov:2006ga}
M.~E. Gusakov and N.~Andersson, ``{Temperature dependent pulsations of
  superfluid neutron stars},'' {\em Mon.Not.Roy.Astron.Soc.} {\bf 372} (2006)
  1776--1790, \href{http://xxx.lanl.gov/abs/astro-ph/0602282}{{\tt
  astro-ph/0602282}}.

\bibitem{Gusakov:2007px}
M.~E. Gusakov, ``{Bulk viscosity of superfluid neutron stars},'' {\em
  Phys.Rev.} {\bf D76} (2007) 083001,
  \href{http://xxx.lanl.gov/abs/arXiv:0704.1071}{{\tt arXiv:0704.1071}}.

\bibitem{Mannarelli:2009ia}
M.~Mannarelli and C.~Manuel, ``{Bulk viscosities of a cold relativistic
  superfluid: Color-flavor locked quark matter},'' {\em Phys.Rev.} {\bf D81}
  (2010) 043002, \href{http://xxx.lanl.gov/abs/0909.4486}{{\tt 0909.4486}}.

\bibitem{Putterman}
S.~J. {Putterman}, ed., {\em {Superfluid hydrodynamics}}, vol.~3, 1974.

\bibitem{Gubser:2008px}
S.~S. Gubser, ``{Breaking an Abelian gauge symmetry near a black hole
  horizon},'' {\em Phys.Rev.} {\bf D78} (2008) 065034,
  \href{http://xxx.lanl.gov/abs/0801.2977}{{\tt 0801.2977}}.

\bibitem{Gubser:2008wv}
S.~S. Gubser and S.~S. Pufu, ``{The Gravity dual of a p-wave superconductor},''
  {\em JHEP} {\bf 0811} (2008) 033,
  \href{http://xxx.lanl.gov/abs/0805.2960}{{\tt 0805.2960}}.

\bibitem{Chen:2010mk}
J.-W. Chen, Y.-J. Kao, D.~Maity, W.-Y. Wen, and C.-P. Yeh, ``{Towards A
  Holographic Model of D-Wave Superconductors},'' {\em Phys.Rev.} {\bf D81}
  (2010) 106008, \href{http://xxx.lanl.gov/abs/1003.2991}{{\tt 1003.2991}}.

\bibitem{Benini:2010pr}
F.~Benini, C.~P. Herzog, R.~Rahman, and A.~Yarom, ``{Gauge gravity duality for
  d-wave superconductors: prospects and challenges},'' {\em JHEP} {\bf 1011}
  (2010) 137, \href{http://xxx.lanl.gov/abs/1007.1981}{{\tt 1007.1981}}.

\bibitem{Yarom:2009uq}
A.~Yarom, ``{Fourth sound of holographic superfluids},'' {\em JHEP} {\bf 0907}
  (2009) 070, \href{http://xxx.lanl.gov/abs/0903.1353}{{\tt 0903.1353}}.

\bibitem{Herzog:2009md}
C.~P. Herzog and A.~Yarom, ``{Sound modes in holographic superfluids},'' {\em
  Phys.Rev.} {\bf D80} (2009) 106002,
  \href{http://xxx.lanl.gov/abs/0906.4810}{{\tt 0906.4810}}.

\bibitem{Amado:2009ts}
I.~Amado, M.~Kaminski, and K.~Landsteiner, ``{Hydrodynamics of Holographic
  Superconductors},'' {\em JHEP} {\bf 0905} (2009) 021,
  \href{http://xxx.lanl.gov/abs/0903.2209}{{\tt 0903.2209}}.

\bibitem{Gubser:2009qf}
S.~S. Gubser and A.~Yarom, ``{Pointlike probes of superstring-theoretic
  superfluids},'' {\em JHEP} {\bf 1003} (2010) 041,
  \href{http://xxx.lanl.gov/abs/0908.1392}{{\tt 0908.1392}}.

\bibitem{Keranen:2009ss}
V.~Keranen, E.~Keski-Vakkuri, S.~Nowling, and K.~Yogendran, ``{Inhomogeneous
  Structures in Holographic Superfluids: I. Dark Solitons},'' {\em Phys.Rev.}
  {\bf D81} (2010) 126011, \href{http://xxx.lanl.gov/abs/0911.1866}{{\tt
  0911.1866}}.

\bibitem{Keranen:2009re}
V.~Keranen, E.~Keski-Vakkuri, S.~Nowling, and K.~Yogendran, ``{Inhomogeneous
  Structures in Holographic Superfluids: II. Vortices},'' {\em Phys.Rev.} {\bf
  D81} (2010) 126012, \href{http://xxx.lanl.gov/abs/0912.4280}{{\tt
  0912.4280}}.

\bibitem{Sonner:2010yx}
J.~Sonner and B.~Withers, ``{A gravity derivation of the Tisza-Landau Model in
  AdS/CFT},'' {\em Phys.Rev.} {\bf D82} (2010) 026001,
  \href{http://xxx.lanl.gov/abs/1004.2707}{{\tt 1004.2707}}.

\bibitem{Arean:2010wu}
D.~Arean, M.~Bertolini, C.~Krishnan, and T.~Prochazka, ``{Type IIB Holographic
  Superfluid Flows},'' \href{http://xxx.lanl.gov/abs/1010.5777}{{\tt
  1010.5777}}.

\bibitem{Keranen:2010sx}
V.~Keranen, E.~Keski-Vakkuri, S.~Nowling, and K.~Yogendran, ``{Solitons as
  Probes of the Structure of Holographic Superfluids},''
  \href{http://xxx.lanl.gov/abs/1012.0190}{{\tt 1012.0190}}.

\bibitem{Herzog:2010vz}
C.~P. Herzog, ``{An Analytic Holographic Superconductor},'' {\em Phys.Rev.}
  {\bf D81} (2010) 126009, \href{http://xxx.lanl.gov/abs/1003.3278}{{\tt
  1003.3278}}.

\bibitem{Minwalla}
J.~Bhattacharya, S.~Bhattacharyya, and S.~Minwalla, ``{to appear},''.

\bibitem{LandL}
L.~D. Landau and E.~M. Lifschitz, {\em {Fluid mechanics}}.
\newblock Course of theoretical physics / by L. D. Landau and E. M. Lifshitz,
  Vol. 6. Elsevier [u.a.], 2~ed., January, 2007.

\bibitem{Son:2005tj}
D.~T. Son, ``{Vanishing bulk viscosities and conformal invariance of unitary
  Fermi gas},'' {\em Phys. Rev. Lett.} {\bf 98} (2007) 020604,
  \href{http://xxx.lanl.gov/abs/cond-mat/0511721}{{\tt cond-mat/0511721}}.

\bibitem{Atkins}
K.~R. Atkins, ``Third and Fourth Sound in Liquid Helium II,'' {\em Phys. Rev.}
  {\bf 113} (Feb, 1959) 962--965.

\bibitem{Hartnoll:2009sz}
S.~A. Hartnoll, ``{Lectures on holographic methods for condensed matter
  physics},'' {\em Class.Quant.Grav.} {\bf 26} (2009) 224002,
  \href{http://xxx.lanl.gov/abs/0903.3246}{{\tt 0903.3246}}.

\bibitem{Herzog:2009xv}
C.~P. Herzog, ``{Lectures on Holographic Superfluidity and
  Superconductivity},'' {\em J.Phys.A} {\bf A42} (2009) 343001,
  \href{http://xxx.lanl.gov/abs/0904.1975}{{\tt 0904.1975}}.

\bibitem{Horowitz:2010gk}
G.~T. Horowitz, ``{Introduction to Holographic Superconductors},''
  \href{http://xxx.lanl.gov/abs/1002.1722}{{\tt 1002.1722}}.

\bibitem{Sachdev:2010ch}
S.~Sachdev, ``{Condensed matter and AdS/CFT},''
  \href{http://xxx.lanl.gov/abs/1002.2947}{{\tt 1002.2947}}.

\bibitem{Balasubramanian:1999re}
V.~Balasubramanian and P.~Kraus, ``{A Stress tensor for Anti-de Sitter
  gravity},'' {\em Commun.Math.Phys.} {\bf 208} (1999) 413--428,
  \href{http://xxx.lanl.gov/abs/hep-th/9902121}{{\tt hep-th/9902121}}.

\bibitem{Bianchi:2001de}
M.~Bianchi, D.~Z. Freedman, and K.~Skenderis, ``{How to go with an RG flow},''
  {\em JHEP} {\bf 0108} (2001) 041,
  \href{http://xxx.lanl.gov/abs/hep-th/0105276}{{\tt hep-th/0105276}}.

\bibitem{Klebanov:1999tb}
I.~R. Klebanov and E.~Witten, ``{AdS / CFT correspondence and symmetry
  breaking},'' {\em Nucl.Phys.} {\bf B556} (1999) 89--114,
  \href{http://xxx.lanl.gov/abs/hep-th/9905104}{{\tt hep-th/9905104}}.

\bibitem{Bhattacharyya:2008xc}
S.~Bhattacharyya, V.~E. Hubeny, R.~Loganayagam, G.~Mandal, S.~Minwalla, {\em
  et.~al.}, ``{Local Fluid Dynamical Entropy from Gravity},'' {\em JHEP} {\bf
  0806} (2008) 055, \href{http://xxx.lanl.gov/abs/0803.2526}{{\tt 0803.2526}}.

\bibitem{Bhattacharyya:2008ji}
S.~Bhattacharyya, R.~Loganayagam, S.~Minwalla, S.~Nampuri, S.~P. Trivedi, {\em
  et.~al.}, ``{Forced Fluid Dynamics from Gravity},'' {\em JHEP} {\bf 0902}
  (2009) 018, \href{http://xxx.lanl.gov/abs/0806.0006}{{\tt 0806.0006}}.

\bibitem{Haack:2008cp}
M.~Haack and A.~Yarom, ``{Nonlinear viscous hydrodynamics in various dimensions
  using AdS/CFT},'' {\em JHEP} {\bf 0810} (2008) 063,
  \href{http://xxx.lanl.gov/abs/0806.4602}{{\tt 0806.4602}}.

\bibitem{Bhattacharyya:2008mz}
S.~Bhattacharyya, R.~Loganayagam, I.~Mandal, S.~Minwalla, and A.~Sharma,
  ``{Conformal Nonlinear Fluid Dynamics from Gravity in Arbitrary
  Dimensions},'' {\em JHEP} {\bf 0812} (2008) 116,
  \href{http://xxx.lanl.gov/abs/0809.4272}{{\tt 0809.4272}}.

\bibitem{Son:2006em}
D.~T. Son and A.~O. Starinets, ``{Hydrodynamics of r-charged black holes},''
  {\em JHEP} {\bf 0603} (2006) 052,
  \href{http://xxx.lanl.gov/abs/hep-th/0601157}{{\tt hep-th/0601157}}.

\bibitem{Natsuume:2010ky}
M.~Natsuume and M.~Ohta, ``{The Shear viscosity of holographic superfluids},''
  \href{http://xxx.lanl.gov/abs/1008.4142}{{\tt 1008.4142}}.

\end{thebibliography}\endgroup

\end{document}